\documentclass[journal ]{new-aiaa}
\usepackage[utf8]{inputenc}
\usepackage{textcomp}
\usepackage{graphicx}
\usepackage{amsmath}
\usepackage[version=4]{mhchem}
\usepackage{siunitx}
\usepackage{longtable,tabularx}
\usepackage{placeins}
\usepackage{booktabs} 
\usepackage{siunitx} 
\usepackage{float}
\usepackage{comment}
\usepackage{makecell}
\usepackage{multirow}
\usepackage{array}
\usepackage{xcolor} 
\usepackage{bm} 

\let\svthefootnote\thefootnote
\newcommand\freefootnote[1]{
  \let\thefootnote\relax
  \footnotetext{#1}
  \let\thefootnote\svthefootnote
}

\usepackage[nolist,nohyperlinks]{acronym}

\title{Distributed Space Resource Logistics Architecture Optimization under Economies of Scale}

\author{Evangelia Gkaravela\footnote{MSc Student, Department of Systems and Enterprises, AIAA Student Member.}}
\affil{Stevens Institute of Technology, Hoboken, NJ, 07030}

\author{Hang Woon Lee\footnote{Assistant Professor, Department of Mechanical, Materials and Aerospace Engineering, AIAA Member.}}
\affil{West Virginia University, Morgantown, WV, 26506}

\author{Hao Chen\footnote{Assistant Professor, Department of Systems and Enterprises; hao.chen@stevens.edu. AIAA Member. (Corresponding Author)}}
\affil{Stevens Institute of Technology, Hoboken, NJ, 07030}

\begin{document}

\freefootnote{This paper is a substantially revised version of the Paper AIAA 2021-4079, presented at the ASCEND 2021 Conference, Las Vegas, Nevada \& Virtual, November 15-17, 2021.}
\maketitle
\begin{acronym}
\acro{ISRU}{In-Situ Resource Utilization}
\acro{GEO}{Geostationary Earth Orbit}
\acro{LEO}{Low Earth Orbit}
\acro{EML1}{Earth-Moon Lagrange Point 1}
\acro{DWE}{Direct Water Electrolysis}
\acro{SWE}{Soil Water Extraction}
\acro{NASA}{National Aeronautics and Space Administration}
\acro{MOXIE}{Mars Oxygen In-Situ Resource Utilization Experiment}
\acro{ACES}{Advanced Cryogenic Evolved Stage}
\acro{MST}{Mission Start Time}
\end{acronym}

\begin{abstract}
This paper proposes an optimization framework for distributed resource logistics system design to support future multimission space exploration. The performance and impact of distributed \ac{ISRU} systems in facilitating space transportation are analyzed. The proposed framework considers technology trade studies, deployment strategy, facility location evaluation, and resource logistics after production for distributed \ac{ISRU} systems. We develop piecewise linear sizing and cost estimation models based on economies of scale that can be easily integrated into network-based mission planning formulations. A case study on a multi-mission cislunar logistics campaign is conducted to demonstrate the value of the proposed method and evaluate key tradeoffs to compare the performance of distributed \ac{ISRU} systems with traditional concentrated \ac{ISRU}. Finally, a comprehensive sensitivity analysis is performed to assess the proposed system under varying conditions, comparing concentrated and distributed \ac{ISRU} systems.

\end{abstract}

\section*{Nomenclature}

{\renewcommand\arraystretch{1.0}
\noindent\begin{longtable*}{@{}l @{\quad=\quad} l@{}}
$\mathcal{A}$  & set of directed arcs \\
$B$  & facility location matrix \\
$\boldsymbol{C}$ & spacecraft capacity vector \\
$\mathcal{C}$ & set of commodities \\
$\mathcal{C_C}$ & set of continuous commodities \\
$\mathcal{C_D}$ & set of discrete commodities \\
 $\bm{c}$   & cost coefficient vector \\
$\bm{d}$ & demand and supply vector \\
$\mathcal{E}$ & set of structures \\
$e$  & structure index \\
$F$  & structure mass \\
$f$  & function intercept \\
$\mathcal{G}$ & network graph \\
$g$  & binary interval variable \\
$H$  & concurrency matrix \\
$h$  & binary interval variable \\
$i$  & node index ($\in \mathcal{N}$) \\
$\mathcal{J}$  & cost \\
$j$  & node index ($\in \mathcal{N}$) \\
$l$  & number of concurrency constraints \\
$\mathcal{N}$ & set of nodes \\
$P$  & flow upper bound \\
$Q$  & transformation matrix \\
$q$ & design quantity \\
$R$ & number of slopes \\
$r$ & index of piecewise function interval \\
$M$ & big constant \\
$M_e$ & bound of piecewise function interval \\
$\mathcal{M}$ & bound of piecewise function interval \\
$\mathcal{T}$ & set of time steps \\
$t$ & time index (integer) \\
$\mathcal{V}$ & set of spacecraft \\
$v$ & spacecraft-type index \\
$W$ & set of time windows \\
$X$ & demanding mass \\
$\bm{x}$ & commodity flow vector \\
$Y$ & manufacturing variable (binary) \\
$\alpha$ & function slope \\
$\beta$ & function slope \\
$\Delta t$ & time of flight \\
\end{longtable*}}

\section{Introduction}
\lettrine{A}{s} low-cost rocket launch technologies mature, large-scale space infrastructure systems become feasible and affordable options to support long-term space exploration campaigns. Among these, \acf{ISRU} systems attract the most attention for their ability to generate resources, especially propellants. Multiple studies on \ac{ISRU} have been conducted, focusing on its chemical processes and standalone system productivity. For example, Schreiner \cite{schreiner2015} established an integrated molten regolith electrolysis \ac{ISRU} model to analyze oxygen generation leveraging melted lunar regolith. Meyen \cite{meyen2017} developed a Mars-atmosphere-based \ac{ISRU} experiment that has been integrated into the Mars 2020 Perseverance Rover, known as \ac{MOXIE}. It, for the first time, produced oxygen on the surface of Mars by solid-oxide electrolysis of atmospheric CO\textsubscript{2}
 in April 2020~\cite{MOXIE_HOFFMAN2023547}. Several testbeds were built by \acs{NASA} \cite{lee2013}, Lockheed Martin \cite{clark2009}, and Orbitec Inc. \cite{gustafson2011} based on hydrogen reduction or carbothermic reduction for oxygen extraction from regolith. These studies demonstrated and evaluated the resource productivity of different chemical processes during space missions through a concentrated \ac{ISRU} system, where all subsystems of an \ac{ISRU} plant are always deployed together.

On the other hand, space logistics research shows the value of the \ac{ISRU} system in supporting long-term space exploration campaigns by generating propellants for spaceflight. Multiple network-based space logistics optimization methods \cite{ishimatsu2016,ho2014,chen2018} demonstrated how \ac{ISRU} infrastructure deployed in earlier missions assists subsequent space transportation. A multi-fidelity optimization framework was proposed to take into account subsystem-level interactions in the \ac{ISRU} system design and technology trade studies \cite{chen2021}. United Launch Alliance also plans to build a cislunar space economy leveraging lunar water \ac{ISRU} for oxygen and hydrogen generation in its Cislunar-1000 program \cite{kutter2016}. These studies only considered the \ac{ISRU} system deployment at a single location and did not fully take advantage of the flexibility of logistics systems to enable distributed \ac{ISRU} system operations.

\begin{figure}[hbt!]
    \centering
    \includegraphics[width=1\textwidth]{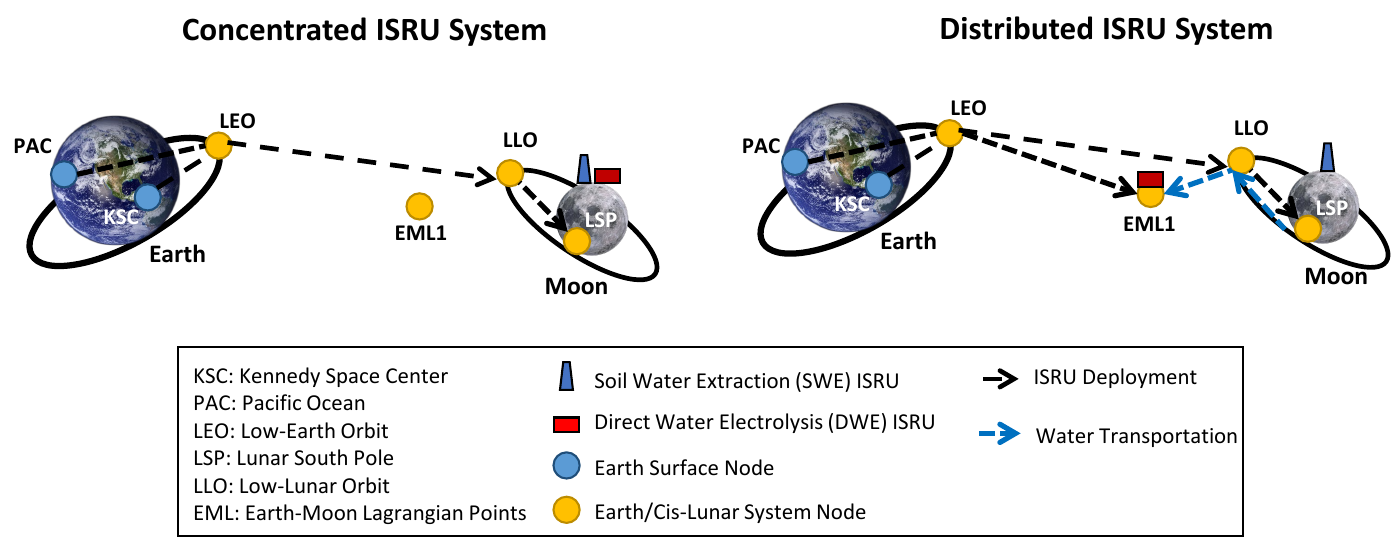}
    \caption{Concentrated ISRU system v.s. distributed ISRU system.}
    \label{Concentrated ISRU System v.s. Distributed ISRU System.}
    \captionsetup{labelfont={color=black}, textfont={color=black}} 
\end{figure}

Past literature focused on analyzing concentrated \ac{ISRU} system performances.  However, this is not always the case for certain types of \ac{ISRU} technologies. For example, for a lunar water \ac{ISRU} plant to generate oxygen and hydrogen based on lunar regolith, two main steps are involved in the operation process. First, water from the lunar regolith is extracted using the \acf{SWE} subsystem, and then the water can be electrolyzed to generate oxygen and hydrogen using the \acf{DWE} subsystem. In fact, only the \ac{SWE} subsystem needs to be deployed on the surface of the moon. The \ac{DWE} subsystem may be deployed on an in-orbit space station between Earth and the moon, such as at Earth-\ac{EML1}. This placement is used as a demonstration case to explore the trade-offs and advantages of utilizing EML1 as a logistics hub for water and propellant transportation. This decision can reduce infrastructure deployment costs and simplify propellant logistics processes after production.
A comparison between concentrated \ac{ISRU} systems and distributed \ac{ISRU} systems is illustrated in Fig. \ref{Concentrated ISRU System v.s. Distributed ISRU System.}. In this scenario, water or ice is transported out of the moon instead of liquid oxygen or hydrogen, which also removes the cryogenic storage system in space vehicles. However, distributed means decentralized material flow management during the system operation. More transportation elements and logistics facilities are involved in the mission operation and system design space. This decentralization introduces another challenge in mission planning, considering the sizing and cost estimation of multiple types of logistics elements simultaneously. Past studies consider these sizing and cost estimation problems either through linear models for both sizing and cost estimation \cite{ishimatsu2016,ho2014,chen2021} or nonlinear sizing models with pre-defined cost metrics \cite{chen2018}. The former methods sacrifice design fidelity to ensure computational efficiency, whereas the latter methods are computationally prohibitive for the consideration of multiple sizing models at the same time and limit the option of cost measures. Previous literature has not effectively addressed the efficient capture of volume discounts for both structure mass and cost estimation in large-scale elements per unit capacity in the context of economies of scale.

In response to this background, this paper proposes a space infrastructure design and deployment optimization framework to model and analyze the impact of distributed systems for future multi-mission space exploration campaigns. Piecewise linear sizing and cost estimation models are proposed based on economies of scale. We integrate these models into network-based mission planning methods as mixed-integer linear programming, whose global optimality is guaranteed.

There are three main contributions achieved in this paper. First, the proposed resource logistics optimization framework captures the economies of scale for both structure sizing and cost estimation in a computationally efficient manner. It enables effective optimization for simultaneously designing multiple transportation elements and plants, which is the main challenge in mission planning with distributed \ac{ISRU}. Nonlinear parametric cost models can be used directly as the mission design metric. Second, the proposed framework can solve facility location and spacecraft manufacturing problems together with transportation mission planning. Traditional space logistics optimization methods focus on transportation flow design. These approaches are sufficient to take into account trade-offs in concentrated \ac{ISRU}. However, facility location problems need to be considered directly in the design space for distributed logistics systems. Finally, to the best of our knowledge, this paper is one of the first studies to evaluate the performance of distributed \ac{ISRU} systems compared with concentrated \ac{ISRU}. This research provides an important step in building a concrete and resilient supply chain for future interplanetary space transportation.

The remainder of this paper is organized as follows: Section \ref{Methodology} introduces the optimization framework for distributed resource logistics systems. It develops sizing and cost estimation models based on economies of scale and integrates them into network-based space logistics optimization methods. A case study on cislunar transportation is conducted in Sec. \ref{Cislunar Transportation Case Study} to compare the performance of a distributed and concentrated lunar water \ac{ISRU} system using the proposed method. Finally, in Sec. \ref{Conclusion}, we conclude this paper and discuss future directions.

\section{Methodology} \label{Methodology}

\subsection{Network-Based Space Logistics Planning Model} \label{Network-Based Space Logistics Planning Model}

Multiple recent studies have shown the effectiveness of converting space mission planning problems into multicommodity network flow problems \cite{ishimatsu2016,ho2014,chen2018,chen2021}. As illustrated in Fig. \ref{Concentrated ISRU System v.s. Distributed ISRU System.} in a network-based logistics model, nodes represent orbits or planets; arcs represent spaceflight trajectories. Then, spacecraft, crew, payload, propellant, etc., are all considered commodities flowing along arcs.


Consider a network graph, $G$, consisting of a set of nodes, denoted by $N$, and a set of arcs, denoted by $\mathcal{A}$. We can write $\mathcal{A} = \{V, N, T\}$ where $V$ is the spacecraft set (index: $v$), $N$ is the node set (index: $i, j$), $T$ is the time step set (index: $t$).

Define the decision variable for commodity flows as $\bm{x}_{vijt}$, which is a continuous variable denoting the amount of commodity flow from node $i$ to node $j$ at time $t$ using spacecraft $v$. 
The transportation mission goals can be represented by logistics supplies and demands, written as $\bm{d}_{it}$ for node $i$ at time $t$. The mission planning objective is to satisfy the demands with the lowest mission cost. As a measure of mission cost, we can define a cost coefficient $\bm{c}_{vij}$. We define a set, $\mathcal{C}$, for commodities considered in the logistics planning. It includes continuous commodity variables, such as propellant, payload, etc., defined by the set $\mathcal{C}_c$, and discrete commodity variables, such as the number of spacecraft or crew members, defined by the set $\mathcal{C}_d$. Then, there are $\mathcal{|C|}$ types of commodities in the logistics system, and $\bm{x}_{vijt}$, $\bm{d}_{it}$, $\bm{c}_{vij}$ are all vectors of dimension $\mathcal{|C|} \times 1$. Typical constraints considered in traditional logistics mission planning methods include the mass balance constraint, the concurrency constraint for flow capacities, and the time-bound constraint for time windows. We define $\Delta t_{ij}$ as the time of flight along the arc from node $i$ to node $j$, $Q_{vij}$ as the transformation matrix for commodity conversion and resource generation during space missions, $H_{vij}$ as the concurrency matrix that enforces commodity flow bounds, such as spacecraft payload capacity and propellant capacity, and finally $W_{ij}$ as the set of time windows for mission operation.

The transformation matrix $Q_{vij}$ is a key component of the optimization framework, capturing the relationships between input and output commodities for spacecraft $v$ as they traverse arcs in the logistics network. Each element of $Q_{vij}$ quantifies how resources are transformed during operations, such as ISRU processes or propellant consumption. The transformation matrix ensures that the conservation of mass and energy principles are maintained across all nodes and arcs, supporting the feasibility of logistics flows.

The concurrency matrix $H_{vij}$ models shared capacity constraints among multiple resources flowing through the same arc. It ensures that the total flow does not exceed the capacity of spacecraft, ISRU plants, or other logistics elements. This matrix is crucial for mixed-resource missions where commodities like propellant and payloads are limited by transportation or processing capacity.

The time windows $W_{ij}$ define the operational periods for nodes or arcs, considering environmental and mission constraints. For example, lunar mining operations may only be active during the lunar day, requiring $W_{ij}$ to capture these periodic availability intervals.

\begin{table}[hbt!]
\centering
\caption{Definitions of indices, variables, and parameters}
\label{table:param}
\begin{tabularx}{\linewidth}{ l X }

\hline
\hline
 Name & \multicolumn{1}{c}{Definition}  \\ 
 \hline
&\multicolumn{1}{c}{\textit{Index}}\\  
\hline
 \(\mathit{v}\)&Spacecraft index \\
 \(\mathit{i,j}\)& Node \\
 \(\mathit{t}\)  & Time step \\
 \(\mathit{l}\)  & Concurrency index \\
 \(\mathit{e}\)  & Structure index \\
 \(\mathit{r}\)  & Index for piecewise sizing function intervals \\
  \(\mathit{\gamma}\)  & Index for piecewise cost estimation function intervals \\
 \hline
&\multicolumn{1}{c}{\textit{Variables}}\\
\hline
\(\bm{x}_\mathit{vijt}\)& Commodity outflows/inflows. Commodities in \(\ \bm{x}^\pm_{\mathit{vijt}} \) are considered as continuous or integer variables based on the commodity type. (\(\mathit{p}\) \(\times\) 1)\\
\(q_{e} \)& Storage capacity for structure e.\\
\(F_{e}^r \)& Mass of structure e.\\
\(P_{\mathit{vijt}}\)& Payload utilization along arc \((i, j)\) at time \(t\). Continuous variable. \\
\(\mathcal{J}_{e} \)& Manufacturing cost of structure e.\\
\(g_{e}^{r} \)& Binary variable for interval r in piecewise function of structure e.\\
\(h_{e}^{\gamma}\)& Binary variable for interval \(\gamma\) in cost estimation function of structure \(e\). \\
\(z_{e}\)& Manufacturing cost of structure \(e\). Continuous variable. \\
\(X_{e}\)& Deployment demand for ISRU plant \(e\). Continuous variable. \\
\(Y_{e}\)& Binary variable indicating if structure \(e\) is manufactured. \\
\(N_{\text{water}} \)& Annual water processing mass for ISRU. \\
\(\mathcal{J}_{\text{Manufacturing}} \)& Total manufacturing cost for logistics elements. Continuous variable. \\
\hline
&\multicolumn{1}{c}{\textit{Parameters}}\\
\hline
\(\bm{c}_\mathit{vijt}\)& Commodity cost coefficient. (\(\mathit{p}\) \(\times\) 1) \\
\(M_{e}^{r}\)& Upper bound of \(r\)-th interval in piecewise sizing function for structure \(e\). \\
\(M_{e}^{\gamma}\)& Upper bound of \(\gamma\)-th interval in cost estimation function for structure \(e\). \\
\(\alpha_{e}^{r}\)& Slope of \(r\)-th interval in piecewise sizing function for structure \(e\). \\
\(\beta_{e}^{\gamma}\)& Slope of \(\gamma\)-th interval in cost estimation function for structure \(e\). \\
\(f_{e}^{r}\)& Intercept of \(r\)-th interval in piecewise sizing function for structure \(e\). \\
\(J_{e}^{\gamma}\)& Intercept of \(\gamma\)-th interval in cost estimation function for structure \(e\). \\
\(\mathit{H}_\mathit{vij}\)&  Concurrency constraint matrix. \((l \times p)\)\\
\(d_{\mathit{it}}\)& Demands or supplies of different commodities at each node. (\(\mathit{p}\) \(\times\) 1)\\
\(\mathit{Q}_\mathit{vij}\)& Commodity transformation matrix. \((\mathit{p}+1) \times (\mathit{p}+1)\)\\
\(\mathit{W}_\mathit{ij}\)& Time window vector. (1 \(\times\) n, where n is the number of time windows)\\
\(\Delta t\) & Time of flight along arc (i,j).\\
\(\beta\)    & Cost estimation function.\\
\hline
\hline
\end{tabularx}
\end{table}

Based on all the notations defined above and at Table \ref{table:param}, the mission planning formulations can be written as follows.

Minimize the objective function:
\begin{equation}
\mathcal{J} = \sum_{(v,i,j,t)\in\mathcal{A}} \bm{c}_{vij}^\top \bm{x}_{vijt}
\label{eq:objective_function}
\end{equation}
 
Subject to

The mass balance constraint:
\begin{equation}
\sum_{(v,j):(v,i,j,t)\in\mathcal{A}} \bm{x}_{vijt} - \sum_{(v,j):(v,j,i,t)\in\mathcal{A}} Q_{vji}\bm{x}_{vji(t-\Delta t_{ji})} \leq \bm{d}_{it} \quad \forall i \in \mathcal{N}, \forall t \in \mathcal{T}
\label{eq:mass_balance}
\end{equation}

The concurrency constraint:
\begin{equation}
H_{vij}\bm{x}_{vijt} \leq \bm{0}_{l \times 1} \quad \forall (v,i,j,t) \in \mathcal{A}
\label{eq:concurrency}
\end{equation}

The time window constraint:
\begin{equation}
\begin{cases}
\bm{x}_{vijt} \geq \bm{0}_{|\mathcal{C}|\times 1}, & \text{if } t \in W_{ij}\\
\bm{x}_{vijt} = \bm{0}_{|\mathcal{C}|\times 1}, & \text{otherwise}
\end{cases} \quad \forall (v,i,j,t) \in \mathcal{A}
\label{eq:time_window}
\end{equation}

\begin{equation*}
\bm{x}_{vijt} = 
\begin{bmatrix}
\bm{x}_{\mathcal{C}} \\
\bm{x}_{D}
\end{bmatrix}_{vijt} \quad \bm{x}_{\mathcal{C}} \in \mathbb{R}^{|\mathcal{C}_c| \times 1}_{\geq 0}, \quad \bm{x}_{D} \in \mathbb{Z}^{|\mathcal{C}_D| \times 1}_{\geq 0}, \quad \forall (v,i,j,t) \in \mathcal{A}
\label{eq:flow_composition}
\end{equation*}

In this formulation, Eq.\eqref{eq:objective_function} is the objective function to minimize the mission cost. Constraint \eqref{eq:mass_balance} is the mass balance constraint to guarantee the commodity outflow is always smaller or equal to the commodity inflow minus the mission demand. The second term of this mass balance constraint describes the commodity transformation along arcs. Constraint \eqref{eq:concurrency} is the concurrency constraint to limit the commodity flow because of the capacities of spacecraft or other facilities. In this constraint, $l$ is the number of capacity types considered in the mission operation. Constraint \eqref{eq:time_window} is the time window constraint. Only when time windows are open are commodity flows permitted. For more detailed settings of each constraint, please refer to Refs.~\cite{chen2018,chen2021}.

This network-based logistics planning method focuses on the flow of commodities during space missions. It enables optimal transportation scheduling and transportation element sizing optimization at the same time. This model is sufficient for concentrated \ac{ISRU} when the facility location is not part of the tradeoff. Studies integrate nonlinear sizing models into the formulation through approximation  \cite{chen2018}. However, the optimization can become easily computationally intractable when multiple element sizing models are considered simultaneously. The linear cost model in the objective function also makes it hard to consider volume discounts in the manufacturing of large space structures. In the following sections, we develop models for economies of scale and manufacturing and facility deployment problems.

\subsection{Economies of Scale Models} \label{Economies of Scale Models}

We care about the mass and cost of building transportation structures with the desired storage capacity or capabilities in space logistics problems. The economies of scale describe the volume discounts that large systems have, such as the lower structure mass or manufacturing cost per unit capacity compared with small structures. An ideal economies of scale model is a concave function where the marginal mass or cost per unit capacity strictly decreases as the structure size increases. 

\begin{figure}[hbt!]
\centering
\includegraphics[width=0.4\textwidth]{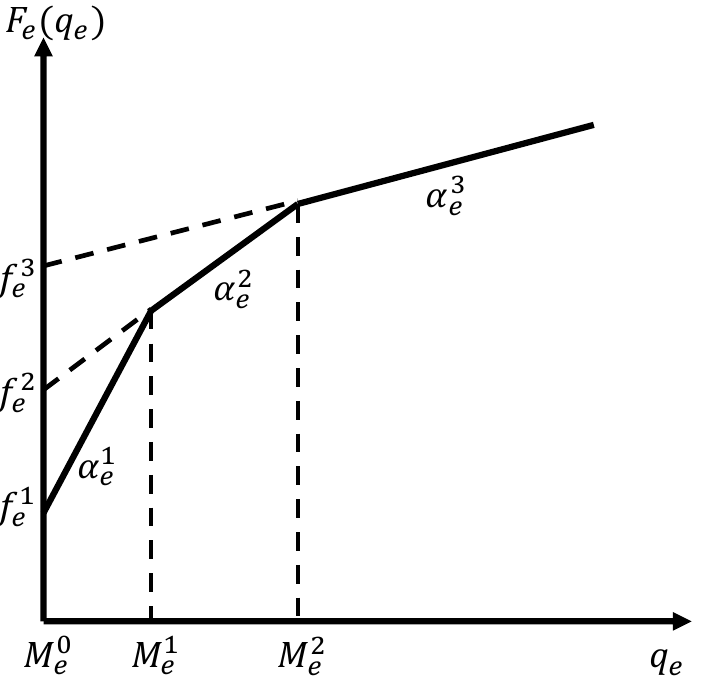}
\caption{Piecewise linear and concave economies of scale model.}
\label{Piecewise Linear and Concave Economies of Scale Model.}
\end{figure}

We define piecewise linear and concave functions for structure sizing and cost estimation, as shown in Fig. \Ref{Piecewise Linear and Concave Economies of Scale Model.}. \cite{magnati,chan,nader} Define $R$ as the number of different slopes in the economies of scale function with index $r = 1,2,..., R$. We use the index $e$ to represent the structure to be designed and defined by the set $\mathcal{E}$, such as spacecraft, \ac{ISRU} plants, or storage tanks. Let $M_{e}^{r-1}$ and $M_{e}^{r}$ be the lower and upper bound of the $r$th interval, respectively, in the piecewise linear function with a slope denoted by $\alpha_{e}^{r}$. Then, we can calculate the function value in each interval using the slope $\alpha_{e}^{r}$ and associated fixed intercept $f_{e}^{r}$. The function, $F_{e}^{r}(q_{e})$, can be expressed as a function of quantity $q_{e}$:

\begin{equation}
F_{e}^{r}(q_{e}) = \alpha_{e}^{r}q_{e} + f_{e}^{r} \quad \forall q_{e} \in (M_{e}^{r-1}, M_{e}^{r}] \quad  \forall r \in [1, \dots, R]
\end{equation}

Note that this concave model assumes that the slopes are strictly decreasing: $\alpha_{e}^{1} > \alpha_{e}^{2} > \ldots > \alpha_{e}^{R}$. However, the function and intercept values are strictly increasing: $0 < f_{e}^{1} < f_{e}^{2} < \ldots < f_{e}^{R}$.

To model the piecewise linear function for economies of scale, a binary interval variable \( g^r_e \) is necessary. This variable indicates whether the quantity \( q_e \) of structure \( e \) lies within the \( r \)th interval of the function, defined by the bounds \( M_e^{r-1} \) (lower bound) and \( M_e^r \) (upper bound). It is expressed as:

\begin{equation}
g_{e}^{r} = 
\begin{cases} 
1, & \text{if } q_{e} \in (M_{e}^{r-1}, M_{e}^{r}] \\
0, & \text{otherwise}
\end{cases}
\label{eq:interval_variable} 
\end{equation}

Then, the quantity \( q_{e} \) can be identified as \( q_{e} = \sum_{r=1}^{R} q_{e}^{r} \) and the following constraints:

\begin{align}
&\sum_{r=1}^{R} g_{e}^{r} \leq 1 \quad \forall e \in \mathcal{E}, \label{eq:sum_ge} \\ 
q_{e}^{r} &\leq M_{e}^{r} g_{e}^{r} \quad \forall e \in \mathcal{E}, \; r = 1, 2, \ldots, R, \label{eq:upper_bound} \\ 
q_{e}^{r} &\geq M_{e}^{r-1} g_{e}^{r} \quad \forall e \in \mathcal{E}, \; r = 1, 2, \ldots, R. \label{eq:lower_bound} 
\end{align}

The binary variable \( g^r_e \) ensures that only one interval is active at a time. The variable \( q^r_e \) represents the portion of the total storage capacity \( q_e \) for structure \( e \) that falls within the \( r \)th interval of the piecewise function.

The function value can be determined using the function:
\begin{equation}
F_{e}^{r}(q_{e}) = \sum_{r=1}^{R} (\alpha_{e}^{r}q_{e}^{r} + g_{e}^{r}f_{e}^{r}) \quad \forall e \in \mathcal{E}
\label{eq:Fe} 
\end{equation}

When the economies of scale need to be considered simultaneously for the structure sizing and cost estimation, we can simply repeat the above steps. For example, if we want to take into account the manufacturing cost in the optimization of the above logistics structure \( e \), then we denote the variable \( q_{e} \) as the storage capacity and the variable \( F_{e} \) as the structure mass, as calculated by Eq.\eqref{eq:Fe} considering constraints Eqs.\eqref{eq:sum_ge}--\eqref{eq:lower_bound}. Here, Eq.\eqref{eq:Fe} is the economies of scale function for structure sizing. The manufacturing cost of structure \( e \), denoted by \( \mathcal{J}_{e} \), is estimated based on the structure mass variable \( F_{e} \). Similarly, we build economies of scale model with independent variable \( F_{e} \), dependent variable \( \mathcal{J}_{e} \), interval index \(\gamma = 1,2,\ldots,\Gamma \), slopes \( \beta_{e}^{\gamma} \), intercepts \( J_{e}^{\gamma} \), and binary interval variable \( h_{e}^{\gamma} \). Here, \(\Gamma \) is the number of linear intervals used to approximate the structure cost function under economies of scale. The cost estimation function for the structure \( e \) can be expressed as:

\begin{equation}
\mathcal{J}_{e}(F_{e}) = \sum_{\gamma=1}^{\Gamma} \left[ \beta^{\gamma}_{e} F^{\gamma}_{e} + h^{\gamma}_{e} J^{\gamma}_{e} \right] \quad \forall e \in \mathcal{E}
\label{eq:Je} 
\end{equation}

Define the lower and upper bounds of \( \gamma \)-th interval of this cost estimation function as \( \mathcal{M}_{e}^{\gamma-1} \) and \( \mathcal{M}_{e}^{\gamma} \). Then, the optimization constraints associated with Eq.\eqref{eq:Je} can be written as:

\begin{align}
&\sum_{\gamma=1}^{\Gamma} h_{e}^{\gamma} \leq 1 \quad \forall e \in \mathcal{E}, \label{eq:sum_he} \\ 
F_{e}^{\gamma} &\leq \mathcal{M}_{e}^{\gamma} h_{e}^{\gamma} \quad \forall e \in \mathcal{E}, \; \gamma = 1, 2, \ldots, \Gamma, \label{eq:upper_bound_he} \\ 
F_{e}^{\gamma} &\geq \mathcal{M}_{e}^{\gamma-1} h_{e}^{\gamma} \quad \forall e \in \mathcal{E}, \; \gamma = 1, 2, \ldots, \Gamma. \label{eq:lower_bound_he} 
\end{align}

\subsection{Manufacturing and Facility Deployment Problems} \label{Manufacturing and Deployment Problems}

The previous section builds the relationship between the structure design quantity \( q_{e} \) and the structure mass \( F_{e} \) and between the mass \( F_{e} \) and the manufacturing cost \( \mathcal{J}_{e} \). Then, there are two problems left for logistics mission planning: 1) whether the structure should be manufactured and 2) how we can integrate the demand in the formulation for facilities to be deployed at a specific location. The structures we discuss here include immovable infrastructure, such as \ac{ISRU} plants, and mobile transportation elements, such as spacecraft and transportation tanks. The only difference between \ac{ISRU} and spacecraft is that an \ac{ISRU} plant should be added to the system at a specific node as a demand, while a spacecraft should be launched from Earth and added to the system as a supply. Therefore, for simplicity, we discuss them using the same set of notations below.

\subsubsection{Manufacturing Problem} \label{Manufacturing Problem}

First, we have to determine whether the structure \( e \) should be manufactured in the logistics system. We define the following binary variable:
\begin{equation}
Y_{e} =
\begin{cases}
1, & \text{if a structure is manufactured} \\
0, & \text{otherwise}
\end{cases}
\quad \forall e \in \mathcal{E}
\label{eq:binary_variable}
\end{equation}

For the purpose of this study, all ISRU plants and spacecraft are assumed to be manufactured on Earth. This assumption reflects the reliance on advanced manufacturing facilities and resources available on Earth.
The challenge of this problem comes from the calculation of the total manufacturing cost in the optimization objective. We know that the manufacturing cost for structure \( e \) is \( \mathcal{J}_{e} \). Then, the total manufacturing cost can be expressed as:
\begin{equation}
\mathcal{J}_{\text{Manufacturing}} = \sum_{e \in \mathcal{E}} Y_{e} \mathcal{J}_{e}
\label{eq:total_manufucturing_cost}
\end{equation}

Because both \( Y_{e} \) and \( \mathcal{J}_{e} \) are decision variables in the optimization problem, Eq. (16) contains quadratic terms, which makes the problem nonlinear. Fortunately, this is a quadratic term of multiplication between a binary variable and a continuous variable. We can define the final manufacturing cost of structure \( e \) as \( z_{e} \). We get:
\vspace{-1ex}
\begin{equation}
z_{e} = Y_{e} \mathcal{J}_{e} \quad \forall e \in \mathcal{E}
\label{eq:structure_manufucturing}
\end{equation}

Using a big constant \( M \), Eq. \ref{eq:structure_manufucturing} can be converted into mixed-integer linear constraints as follows:
\begin{align}
z_{e} &\leq M Y_{e} \quad \forall e \in \mathcal{E} \label{eq:z_upper}\\
z_{e} &\leq \mathcal{J}_{e} \quad \forall e \in \mathcal{E} \label{eq:z_cost}\\
z_{e} &\geq \mathcal{J}_{e} - (1 - Y_{e})M \quad \forall e \in \mathcal{E} \label{eq:z_lower}\\
z_{e} &\geq 0 \quad \forall e \in \mathcal{E} \label{eq:z_nonnegative}
\end{align}

\subsubsection{Facility Deployment Problem} \label{Deployment Problem}

After determining the structure to be manufactured for the logistics, we need to generate demand for the \ac{ISRU} plant deployment or the spacecraft supply. Now, we consider the deployment of \ac{ISRU} plants and spacecraft separately because the demand for \ac{ISRU} plants is a continuous variable, while the supply of spacecraft is a discrete variable. We can represent the set of structures as \(\mathcal{E} = \left[ \mathcal{E}_{\text{ISRU}}, \mathcal{E}_{\text{SC}} \right]^{\top}\), where \(\mathcal{E}_{\text{ISRU}}\) denotes the set of \ac{ISRU} plants and \(\mathcal{E}_{\text{SC}}\) denotes the set of spacecraft or transportation tanks.

For \ac{ISRU} deployment demand, define the demanding mass for ISRU plant \( e \) as:
\begin{equation}
X_e = Y_e F_e \quad \forall e \in \mathcal{E}_{\text{ISRU}}
\label{eq:demaning_mass_isru} \\
\end{equation}
Then, we define a binary facility location matrix \( B_{iet} \) where \( i \) is the node index of the \ac{ISRU} facility deployment site. The binary elements of \( B_{iet} \) are determined in advance depending on the \ac{ISRU} operation requirement and landing site environment. Specifically,  $B_{iet}=1$ indicates that \acs{ISRU} structure $e$ is deployed at node i at time step t, while $B_{iet}=0$ indicates that the structure is not deployed.  We can get the total \ac{ISRU} demand mass to be deployed at node \( i \) at a specific time \( t \) as:
\vspace{-1ex}
\begin{equation}
d_{\text{ISRU},it} = -\sum_{e \in \mathcal{E}_{\text{ISRU}}} B_{iet} X_e \quad \forall i \in \mathcal{N} \quad 
 \forall t \in \mathcal{T}
\label{eq:total_demand_mass} 
\end{equation}

Note that based on the definition of the mass balance constraint Eq.\eqref{eq:mass_balance}, all supplies are positive and demands are negative in the demand vector $\mathbf{d}_{it}$. In Eq.\eqref{eq:total_demand_mass}, \( X_e \) is a product of a binary variable and a continuous variable. Similarly, Eq.\eqref{eq:total_demand_mass} can be linearized by following mixed-integer linear constraints:
\vspace{-1ex}
\begin{align}
X_e &\leq M Y_e \quad
\quad \forall e \in \mathcal{E}_{\text{ISRU}} \\
X_e &\leq F_e \quad \forall e \in \mathcal{E}_{\text{ISRU}} \\
X_e &\geq F_e - (1 - Y_e)M \quad \forall e \in \mathcal{E}_{\text{ISRU}} \\
X_e &\geq 0 \quad \forall e \in \mathcal{E}_{\text{ISRU}}
\end{align}

For spacecraft supply, because all spacecraft need to be launched from Earth, the spacecraft supply can be expressed as: 

\vspace{-1ex}
\begin{equation}
d_{\text{SC},i} =
\begin{cases}
\sum\limits_{e \in \mathcal{E}_{\text{SC}}} Y_{e}, & \text{if } i = \text{Earth} \\
0, & \text{otherwise}
\end{cases}
\label{eq:spacecraft_supply}
\end{equation}

We can also include additional demands for spacecraft if we want to specify the target orbit of spacecraft launching.

\subsection{Distributed Logistics System Optimization Framework}

In this section, we integrate the economies of scale model proposed in \ref{Economies of Scale Models} and the manufacturing and deployment formulation proposed in \ref{Manufacturing and Deployment Problems} into the network-based logistics planning model. Based on the notations mentioned above, the optimization framework for the distributed logistics system can be written as follows:


\begin{equation}
\textbf{Min} \quad 
\mathcal{J} = \sum_{(v,i,j,t) \in \mathcal{A}} \mathbf{c}_{vij}^\top \bm{x}_{vijt} + \sum_{e \in \mathcal{E}} z_e
\label{eq:objective_function_distributed} 
\end{equation}
Subject to:

\textit{section 1 (space logistics flows)}
\vspace{-3ex}
\begin{equation}
\sum_{(v,j):(v,i,j,t) \in \mathcal{A}} \bm{x}_{vijt} - \sum_{(v,j):(v,j,i,t) \in \mathcal{A}} Q_{vji} \bm{x}_{vji(t - \Delta t_{ji})} \leq 
\label{} 
\begin{bmatrix}
\bm{d}_{i} \\
d_{\text{ISRU},i} \\
d_{\text{SC},i}
\end{bmatrix}_t 
\quad \forall i \in \mathcal{N} \quad \forall t \in \mathcal{T}
\label{eq:mass_balance_constraint_distributed}
\end{equation}

\vspace{-3ex}
\begin{equation}
H_{vij} \bm{x}_{vijt} \leq \left( \sum_{r=1}^{R} \bm{q}^{r}_{\mathrm{sc}} \right) x_{\mathrm{sc},vijt} \quad \forall (v,i,j,t) \in \mathcal{A}
\label{eq:concurrency_constraint_distributed}
\end{equation}

\vspace{-3ex}
\begin{equation}
\begin{cases}
\bm{x}_{vijt} \geq \bm{0}_{|\mathcal{C}| \times 1} & \text{if } t \in W_{ij}, \\
\bm{x}_{vijt} = \bm{0}_{|\mathcal{C}| \times 1} & \text{otherwise}
\end{cases} 
\quad \forall (v,i,j,t) \in \mathcal{A}
\label{eq:time_window_constraint_distributed}
\end{equation}

\makebox[0.7\textwidth][c]{\textit{section 2 (sizing model)} \hfill Eqs. \eqref{eq:sum_ge}-\eqref{eq:Fe}}

\vspace{-1ex}
\makebox[0.7\textwidth][c]{\textit{section 3 (cost model)} \hfill Eqs.\eqref{eq:Je}-\eqref{eq:lower_bound_he}}

\makebox[0.7\textwidth][c]{\textit{section 4 (manufacturing cost)} \hfill Eqs.\eqref{eq:z_upper}-\eqref{eq:z_nonnegative}}

\makebox[0.7\textwidth][c]{\textit{section 5 (demand generation)} \hfill Eqs.\eqref{eq:total_demand_mass}-\eqref{eq:spacecraft_supply}}

\begin{equation*}
\bm{x}_{vijt} = 
\begin{bmatrix}
\bm{x}_{C} \\
\bm{x}_{D}
\end{bmatrix}_{vijt}, \quad \bm{x}_{C} \in \mathbb{R}_{\geq 0}^{|C_c| \times 1}, \quad \bm{x}_{D} \in \mathbb{Z}_{\geq 0}^{|\mathcal{C_D}| \times 1}, \quad \forall (v,i,j,t) \in \mathcal{A}
\label{eq:flow_composition_distributed}
\end{equation*}

\begin{equation*}
g_e^r, h_e^{\gamma} \in \{0, 1\} \quad \forall e \in \mathcal{E}, \forall r, \gamma \quad q_e^r, F_e^{\gamma}, z_e, X_e \in \mathbb{R}_{\geq 0} \quad \forall e \in \mathcal{E}, \forall r, \gamma
\label{eq:variable_definitions}
\end{equation*}

In this formulation, Eq. \eqref{eq:objective_function_distributed} is the objective function to minimize the total mission cost. The first term covers all transportation costs determined by commodity flows, including the facility deployment cost. The second term covers all manufacturing costs of \ac{ISRU} plants and transportation elements. The objective function, Eq. \eqref{eq:objective_function_distributed} combines the transportation cost, incurred incrementally during the mission, and the non-recurring plant manufacturing cost. While this formulation can be used for lifecycle cost estimation, its validity depends on mission duration, plant type, and reuse potential. If the ISRU plant is reusable across missions, its manufacturing cost should be spread across its lifecycle. In this research, we focus on cost estimation during the mission time span and assume the deployed ISRU will only be used for the proposed mission scenarios.

Additional demand and supply generated by the \ac{ISRU} and spacecraft deployment are added to the demand vector of the mass balance constraint, Eq.\eqref{eq:mass_balance_constraint_distributed}. Equation \eqref{eq:concurrency_constraint_distributed} is the concurrency constraint that limits the commodity flow because of the capacity of structures. We can consider the payload and propellant capacities from spacecraft, transportation tanks, or \ac{ISRU} storage systems. The vector variable \( \bm{q}^{r} \) has a dimension of \( |\mathcal{X}| \times 1 \). It is the vector form of the capacity of commodities. The flow variable \( x_{sc,ijt} \) is a binary variable defined specifically for the spacecraft \( v \). When we consider spacecraft sizing, the spacecraft set \( \mathcal{V}\) is equal to \( \mathcal{E}_{SC} \). The expression \( \sum_{r=1}^{R} \bm{q}^{r} \) determines the capacity of the spacecraft \( v \) during spaceflight. We can define a \( |\mathcal{V}| \times 1 \) capacity vector variable \( \bm{C} \) for all spacecraft in the set \( \mathcal{V}\).


\begin{equation}
\bm{C} = \sum_{r=1}^{R} \bm{q}^{r}
\end{equation}

Then, the right-hand side of the concurrency constraint Eq.\eqref{eq:concurrency_constraint_distributed} can be expressed as \(P_{vijt} = C_{v}x_{sc,vijt}\). This is also a product of a binary variable and a continuous variable. We can express this term equivalently using the following mixed-integer programming formulations:
\vspace{-1ex}
\begin{align}
P_{vijt} &\leq M x_{sc,vijt} \quad \forall (v, i, j, t) \in \mathcal{A} \label{eq:capacity_constraint_upper}\\
P_{vijt} &\leq C_{v} \quad \forall (v, i, j, t) \in \mathcal{A} \label{eq:spacecraft_capacity_limit}\\
P_{vijt} &\geq C_{v} - (1 - x_{sc,vijt})M \quad \forall (v, i, j, t) \in \mathcal{A} \label{eq:spacecraft_utilization}\\
P_{vijt} &\geq 0 \quad \forall (v, i, j, t) \in \mathcal{A} \label{eq:non_negativity_payload}
\end{align}
When spacecraft or transportation tank sizing problem is considered, Eqs.\eqref{eq:capacity_constraint_upper}-\eqref{eq:non_negativity_payload} need to be added into the optimization framework to replace the right-hand side of Eq.\eqref{eq:concurrency_constraint_distributed} by \(P_{vijt}\). A similar technique was used in Ref. \cite{chen2018} to handle the multiplication of spacecraft capacity and spacecraft flow variables.

\section{Cislunar Transportation Case Study} \label{Cislunar Transportation Case Study}
In this section, we develop a case study on cislunar logistics to compare the performance of distributed \ac{ISRU} and concentrated \ac{ISRU} systems using the proposed optimization framework. We first introduce mission operation assumptions and economies of scale models used in the case study in Section \Ref{Mission Scenarios}. Then, we show mission planning results and discuss mission performances in \Ref{Results and Analysis}.

\subsection{Mission Scenarios} \label{Mission Scenarios}

\begin{figure}[hbt!]
\centering
\includegraphics[width=0.7\textwidth]{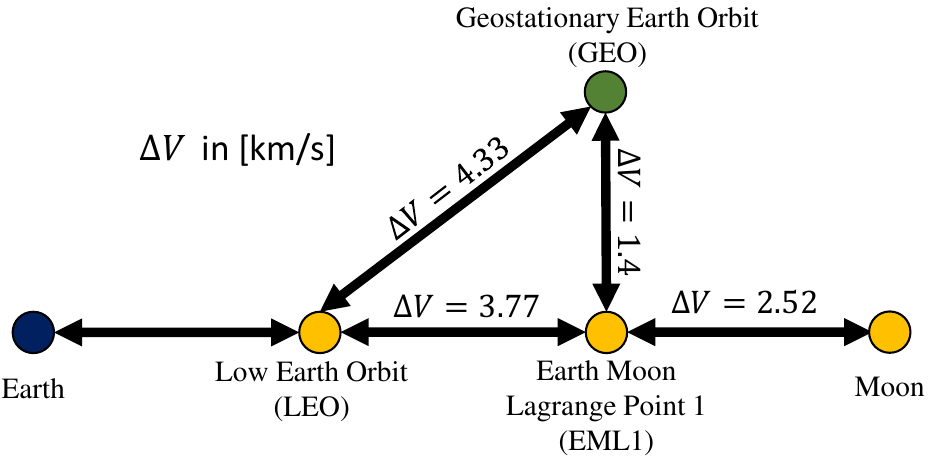}
\caption{Cislunar transportation network model.}
\label{fig:cislunar_network}
\end{figure}

The mission planning scenario considered in this paper is a series of transportation missions in the cislunar system. The transportation network model is shown in Fig. \ref{fig:cislunar_network}. This transportation network includes Earth, \ac{LEO}, \ac{GEO}, \ac{EML1}, and the moon. The trajectory $\Delta V$ along each arc is also shown in Fig. \ref{fig:cislunar_network}. In this case study, we assume that the time of flight along each arc is always one-time step. The mission goals are to deliver 25,000 kg payload to \ac{GEO} and 15,000 kg payload to the moon from Earth every year. At the same time, the logistics system needs to satisfy 5,000 kg oxygen annual demands at \ac{EML1} and \ac{GEO} that can be used for astronaut activities or as the propellant oxidizer. To structure the timeline of these mission activities, we define the \ac{MST} as the initial reference point from which all mission operations are measured. All subsequent time steps, such as annual demands and mission events, are defined relative to \ac{MST}. This framework allows for a consistent comparison of different \ac{ISRU} strategies, including their setup phases and operational efficiencies.

The oxygen can be delivered from Earth, generated by the lunar \ac{ISRU} system on the Moon, or generated by a distributed \ac{ISRU} system on-site. This case study aims to analyze the ability of \ac{ISRU} to support transportation logistics in the cislunar system and compare the performance of concentrated and distributed \ac{ISRU} systems. The mission demands and supplies are summarized in Table \Ref{tab:mission-demands-supplies}. All ISRU plants are assumed to be manufactured on Earth, similar to spacecraft. Also, all resources are supplied from Earth in infinite quantities. However, we need to pay corresponding costs for the rocket launch, spaceflight operation, and system manufacturing to use these resources.

\begin{table}[hbt!]
\centering
\caption{Mission Demands and Supplies.}
\label{tab:mission-demands-supplies}
\begin{tabular}{lccc}
\hline
\textbf{Payload Type}        & \textbf{Node} &  \textbf{Time, [days]} & \makecell{\textbf{Demand (-)}/\textbf{Supply(+)}, \\ \textbf{[kg]}} \\
\hline
Payload             & \ac{GEO}  & Annually         & -25,000    \\
Oxygen              & \ac{GEO}  & Annually         & -5,000     \\
Oxygen              & \ac{EML1} & Annually         & -5,000     \\
Payload             & Moon & Annually         & -15,000    \\
\multirow{2}{*}{\makecell[{{l}}]{Propellant, ISRU \\ Maintenance, spares}} & 
\multirow{2}{*}{Earth} & 
\multirow{2}{*}{All the time} & 
\multirow{2}{*}{$+\infty$} \\
 & & & \\
\hline
\end{tabular}
\end{table}

In this case study, we focus on the analysis of lunar water \ac{ISRU}. For simplicity, we assume the \ac{ISRU} system is mainly made up of two components: 1) The \ac{SWE} system to extract water from lunar soil; 2) The \ac{DWE} system to electrolyze water into oxygen and hydrogen. As we can see, depending on the working environment, the \ac{SWE} system needs to be deployed on the lunar surface. In contrast, the \ac{DWE} system can be deployed either together with \ac{SWE} or on-site at the place of resource demand. In this mission scenario, the problem is whether we should deploy a concentrated \ac{ISRU} on the moon or deploy a distributed \ac{ISRU} system with the \ac{SWE} system on the moon and the \ac{DWE} system on \ac{EML1}. Based on the \ac{ISRU} subsystem-level modeling established in Ref. \cite{chen2020}, for 5.6\% water concentration lunar regolith, a 1 kg \ac{SWE} reactor can extract 31.4 kg water per year, and a 1 kg \ac{DWE} reactor can electrolyze 105.2 kg water per year. We assume a continuous operation of \ac{ISRU} systems during the space mission. The \ac{ISRU} system also contains multiple subsystems to support the operation of the reactors, such as the power system, storage system, soil extraction system, etc. In this analysis, we assume these subsystems are integrated into the \ac{SWE} and \ac{DWE} systems. As a result, we added a 200\% overhead mass for these subsystems as the baseline \ac{ISRU} productivity. Therefore, 1 kg of \ac{SWE} reactor can extract 10.5 kg of water per year, and 1 kg of \ac{DWE} system can electrolyze 35 kg of water per year. For the economies of scale model, the \ac{ISRU} plant mass is a concave function of the \ac{ISRU} productivity. However, in logistics mission planning, \ac{ISRU} structure is part of the commodity flow variables. Therefore, we switch the x and y axes to create a model where the \ac{ISRU} productivity is a function of the \ac{ISRU} structure mass. The economies of scale model derivation is still valid. With the increase of system mass, we assume the \ac{ISRU} productivity will increase 10\% per 3000 kg of structure mass. The baseline manufacturing cost is assumed as the same for both the \ac{SWE} system and the \ac{DWE} system based on the structure mass, \$10,000/kg \cite{chen2020}. As the increase of the structure mass, the \ac{ISRU} system manufacturing cost per unit mass will decrease 10\% per 3000 kg \cite{tribe, ouyang}. The sizing and cost models of \ac{ISRU} are shown as follows:

\vspace{-0.3 cm} 

\begin{align}
N_{\text{water}}(F_{\text{SWE}}) &= 
\begin{cases}
       10.5F_{\text{SWE}} & \forall F_{\text{SWE}} \in (0,3000] \\
    N_{\text{water}}(3000) + 10.5 \times (1 + 10\%)(F_{\text{SWE}} - 3000) & \forall F_{\text{SWE}} \in (3000,6000] \\
    N_{\text{water}}(6000) + 10.5 \times (1 + 10\%)^2(F_{\text{SWE}} - 6000) & \forall F_{\text{SWE}} \in (6000,9000] \\ &\vdots
\end{cases} \tag{38} \label{eq:waterisru} \\ 
N_{\text{water}}(F_{\text{DWE}}) &= 
\begin{cases}
      35F_{\text{DWE}} & \forall F_{\text{DWE}} \in (0,3000] \\
    N_{\text{water}}(3000) + 35 \times (1 + 10\%)(F_{\text{DWE}} - 3000) & \forall F_{\text{DWE}} \in (3000,6000] \\
    N_{\text{water}}(6000) + 35 \times (1 + 10\%)^2(F_{\text{DWE}} - 6000) & \forall F_{\text{DWE}} \in (6000,9000] \\ & \vdots
\end{cases} \tag{39} \\
\mathcal{J}_{\text{ISRU}}(F_{\text{ISRU}}) &= 
\begin{cases}
   \mathcal{J}_{\text{ISRU}}(1000) & \forall F_{\text{ISRU}} \in (0,1000] \\ 
    \mathcal{J}_{\text{ISRU}}(1000)+10000(F_{\text{ISRU}}-1000)  & \forall F_{\text{ISRU}} \in (1000,3000] \\
    \mathcal{J}_{\text{ISRU}}(3000) + 10000 \times (1 - 10\%)(F_{\text{ISRU}} - 3000) & \forall F_{\text{ISRU}} \in (3000,6000] \\
    \mathcal{J}_{\text{ISRU}}(6000) + 10000 \times (1 - 10\%)^2(F_{\text{ISRU}} - 6000) & \forall F_{\text{ISRU}} \in (6000,9000] \\ &\vdots
\end{cases} \tag{40} \label{eq:isru}
\end{align}

\begin{figure}[hbt!]
\centering
\includegraphics[width=\textwidth]{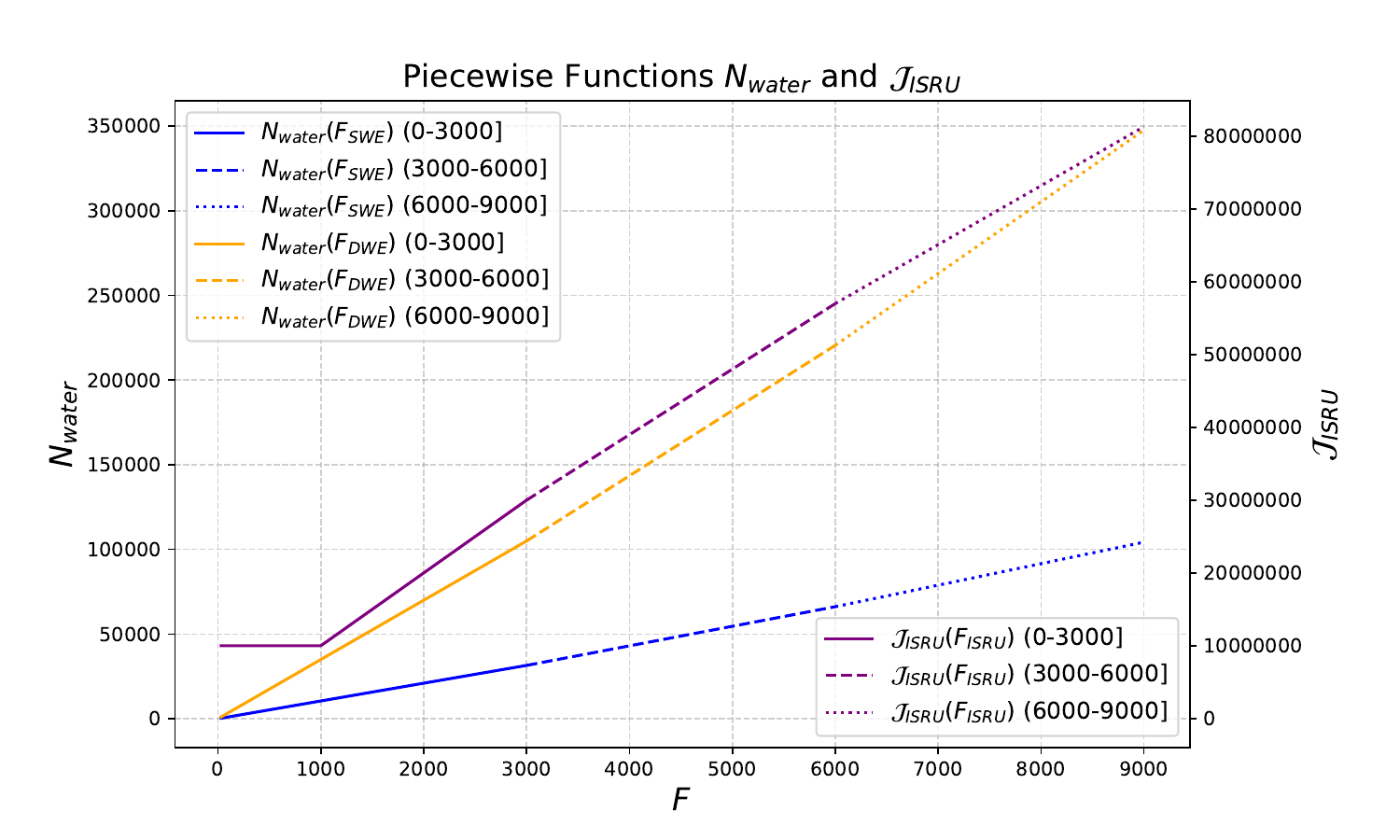}
\caption{Sizing and cost models of ISRU.}
\label{fig:cost_models}
\end{figure}

The nonlinear functions in Eqs. \eqref{eq:waterisru}-\eqref{eq:isru} represent piecewise-defined economies of scale relationships, where each interval accounts for different structural efficiencies and cost scaling factors. Equations \eqref{eq:waterisru}-\eqref{eq:isru} are depicted in Fig. \ref{fig:cost_models}. In these functions, $N_{\text{water}}$ is the annual water processing mass of the \ac{ISRU} system, $F_{\text{ISRU}}$ is the structure mass for \ac{SWE} and \ac{DWE} systems, and $\mathcal{J}_{\text{ISRU}}$ is the \ac{ISRU} manufacturing cost. The maintenance of the \ac{ISRU} system is also considered. It is assumed that the maintenance spares required are equal to 5\% of the \ac{ISRU} structure mass, and the manufacturing cost of maintenance spares is always \$10,000/kg.

After resources are generated, we need spacecraft to deliver them to the demanding node. For the simplicity of the optimization, we do not consider the concurrent sizing of spacecraft in this paper. Instead, we use spacecraft with a fixed design. Based on the design of \ac{ACES} \cite{kutter2016} spacecraft and Centaur \cite{dekruif2007} spacecraft, the propulsive stage of this spacecraft uses LO\textsubscript{2}/LH\textsubscript{2} as the propellant. Therefore, the oxygen and hydrogen generated by the \ac{ISRU} can also support the spaceflight of the spacecraft. The propellant mass ratio of oxygen and hydrogen is 5.5:1. The structure mass of the spacecraft is assumed as 6,000 kg, and it has a propellant capacity of 65,000 kg. The manufacturing cost of one spacecraft is \$150M, and the operation cost per spaceflight is \$0.5M. We assume that there are two spacecraft available on the Earth at the beginning of each time window. We assume that water and LO\textsubscript{2}/LH\textsubscript{2} propellant are both delivered and stored through specific water and propellant tanks. The water tank capacity ratio is assumed as 40 kg of \ce{H2O} per kg of tank mass and its cost is \$800/kg. However, the transportation of oxygen and hydrogen requires cryogenic cooling systems to prevent boil-off. The LO\textsubscript{2}/LH\textsubscript{2} propellant tank capacity ratio is assumed as 1.478 kg of LO\textsubscript{2}/LH\textsubscript{2} per kg of tank and its cost is \$1,869/kg. The assumptions for mission operation are summarized in Table \ref{tab:mission-operation}.

 
\begin{table}[htb!]
\centering
\caption{Assumptions of mission operation.}
\label{tab:mission-operation}
\begin{tabular}{@{}lll@{}}
\toprule
\multicolumn{1}{l}{} & \textbf{Parameter} & \textbf{Assumed Value} \cite{chen2020,kutter2016}\\
\midrule
\multirow{7}{*}{\textbf{Mission Operations}} 
& Propellant & LO\textsubscript{2}/LH\textsubscript{2}
 \\
& $I_{sp}$ & 420 s \\
& Time window & Every 6 months \\
& Spacecraft propellant capacity & 65,000 kg \\
& Spacecraft structure mass & 6,000 kg \\
& Available spacecraft each time window & 2 spacecraft \\
&Water Tank Capacity Ratio     & 40 kg of H\textsubscript{2}O capacity per kg of tank 
\\
&Propellant Tank Capacity Ratio     & 1.478 kg of LO\textsubscript{2}/LH\textsubscript{2} per kg of tank \\
& \ac{ISRU} maintenance & 5\% plant mass/year  \\
\midrule
\multirow{7}{*}{\textbf{Cost Models}} 
& Rocket launch cost & \$5,000/kg \\
& Spacecraft manufacturing cost & \$150M each \\
& Spaceflight operation cost & \$0.5M/flight \\
&Water Tank Cost       &\$800/kg   \\
&Propellant Tank Cost    & \$1,869/kg \\
& LO\textsubscript{2}
 cost on Earth & \$0.15/kg \\
& LH\textsubscript{2}
 cost on Earth & \$5.97/kg \\
& \ac{ISRU} maintenance spares cost & \$10,000/kg \\
\bottomrule
\end{tabular}
\end{table}

\subsection{Results and Analysis} \label{Results and Analysis}

\subsubsection{Mission Planning Results} \label{Mission Planning Results}

We perform mission planning to satisfy the transportation mission demands defined in Table \ref{tab:mission-demands-supplies} for three consecutive years. To compare the performances of concentrated and distributed \ac{ISRU} systems during their steady-state operations, we assume the deployment of an initial 1000 kg \ac{ISRU} plant at the start of the mission. This initial deployment aims to mitigate the impact of the setup phase, particularly for distributed \ac{ISRU} systems, which require longer times to become fully operational. 

As seen in Table \ref{tab:comparison-isru-systems}, the total mission cost using concentrated \ac{ISRU} is \$2.922B, and an 11,028 kg \ac{ISRU} system is deployed in total. In comparison, the total mission cost using distributed \ac{ISRU} is \$2.916B, and a 11,601 kg \ac{ISRU} system is deployed.

\begin{figure}[H]
\centering
\includegraphics[width=\textwidth]{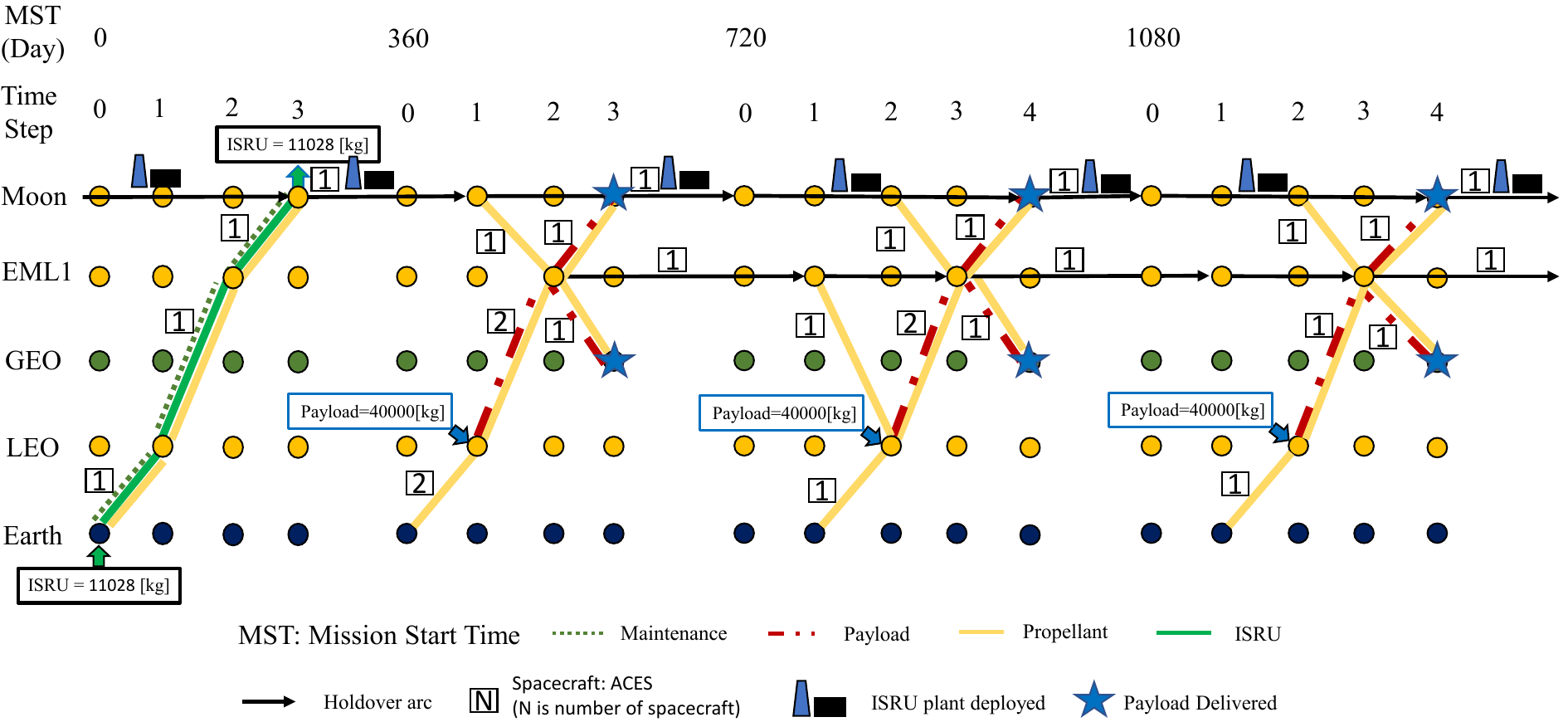}
\caption{Mission planning using concentrated ISRU.}
\label{fig:concentrated_ISRU}
\end{figure}

\begin{figure}[H]
\centering
\includegraphics[width=\textwidth]{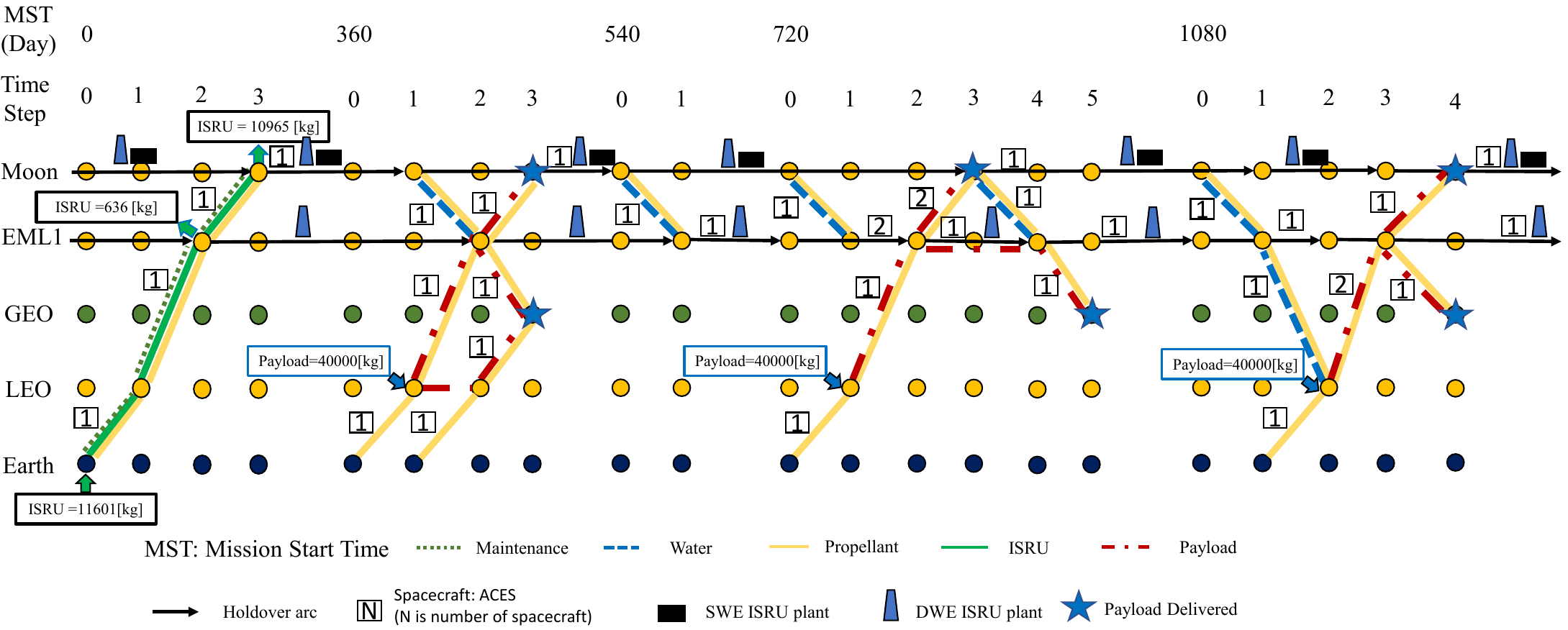}
\caption{Mission planning using distributed ISRU.}
\label{fig:distributed_ISRU}
\end{figure}

Figures \ref{fig:concentrated_ISRU} and \ref{fig:distributed_ISRU} illustrate detailed mission planning solutions using concentrated and distributed ISRU systems. Comparing these two figures, first, we can find that for the deployment demand at \ac{GEO}, the logistics system chooses a LEO-EML1-GEO logistics path instead of a LEO-GEO logistics path. This is because spacecraft can get refueled at \ac{EML1} from the lunar \ac{ISRU}, which makes the spaceflight from \ac{EML1} to \ac{GEO} get rid of relying on the propellant launched from Earth. However, as shown in Fig. \ref{fig:distributed_ISRU}, part of the payload is still deployed through the LEO-GEO path. This is caused by the long setup phase of the distributed \ac{ISRU} system. In this mission scenario, the \ac{SWE} system needs to extract water from the lunar surface. Then, part of the water is electrolyzed by the \ac{DWE} plant deployed on the moon, and the remaining water needs to be transported to the \ac{DWE} plant deployed on \ac{EML1} to generate propellant. The lead time of water logistics delays the propellant generation, which elongates the setup of the distributed \ac{ISRU} system.

\subsubsection{Impact of a Setup Phase}

\begin{figure}[hbt!]
\centering
\includegraphics[width=\textwidth]{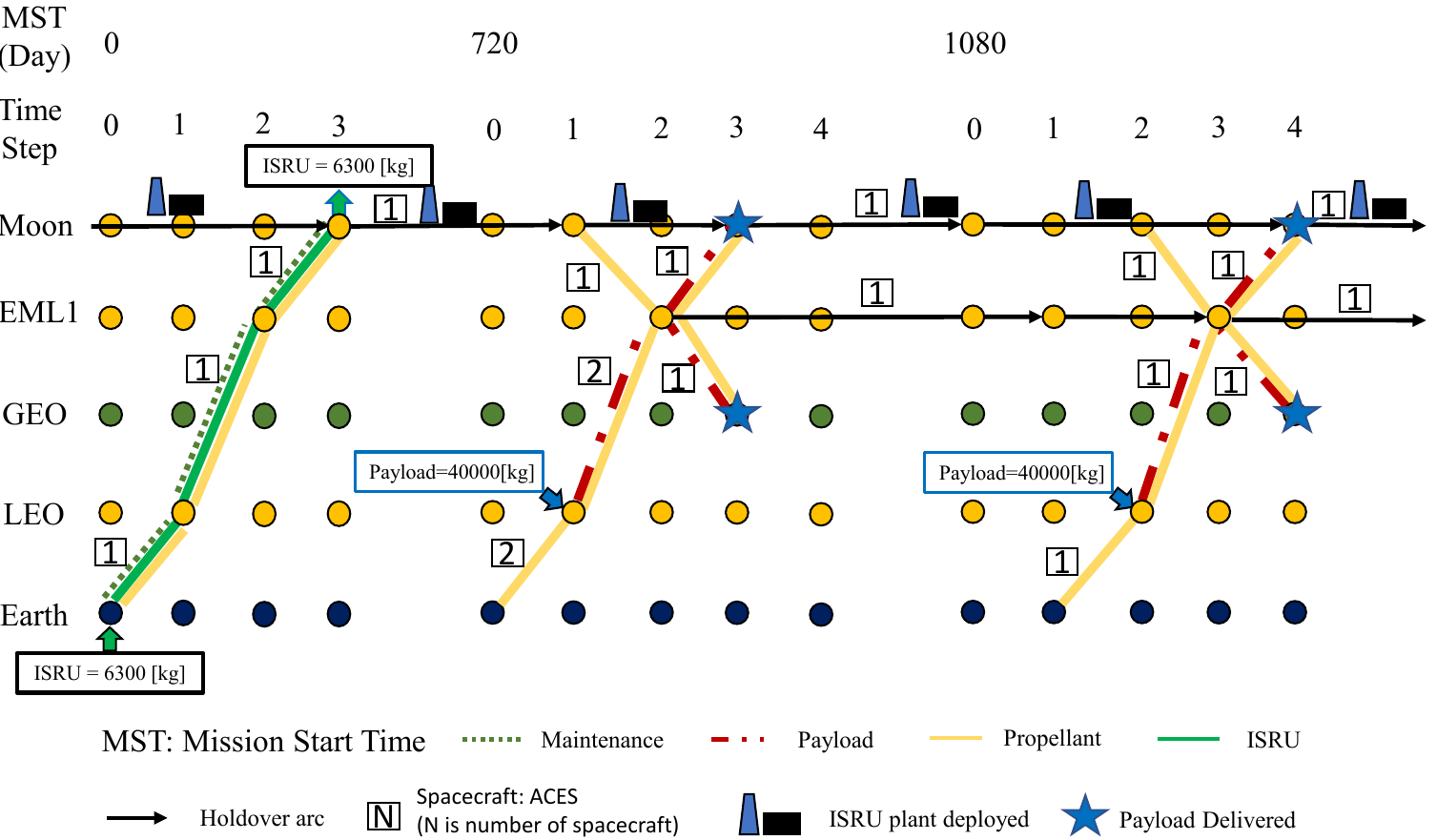}
\caption{Mission planning using concentrated \ac{ISRU} with setup phase.}
\label{fig:concentrated_ISRU_setup}
\end{figure}

\begin{figure}[hbt!]
\centering
\includegraphics[width=\textwidth]{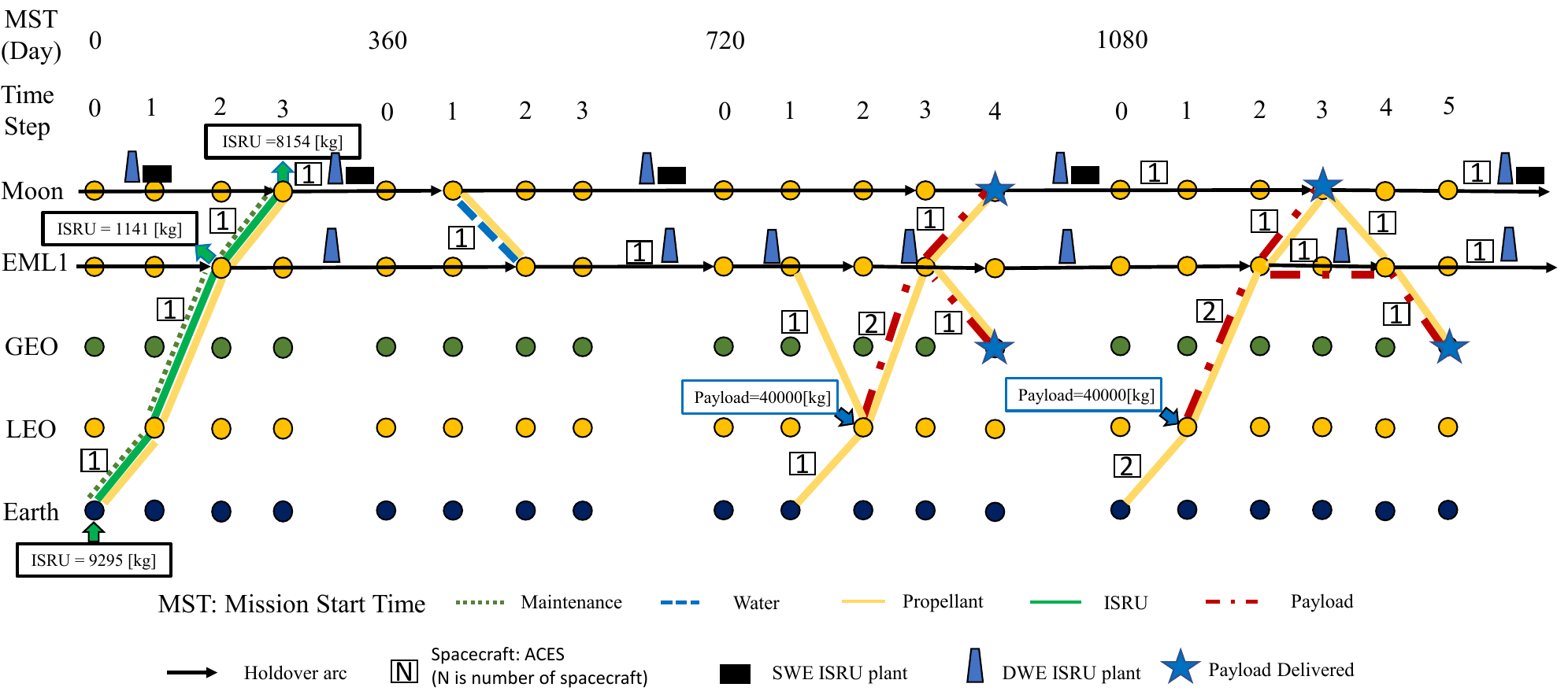}
\caption{Mission planning using distributed ISRU with setup phase.}
\label{fig:distributed_ISRU_setup}
\end{figure}

To evaluate the impact of the \ac{ISRU} system setup phase, we consider mission planning with a one-year setup phase where mission demands are satisfied starting from the second mission. The mission planning solutions using concentrated and distributed \ac{ISRU} systems are shown in Fig. \ref{fig:concentrated_ISRU_setup} and Fig. \ref{fig:distributed_ISRU_setup}, respectively. When a long setup phase is considered at the beginning of the mission, the total mission cost using a distributed \ac{ISRU} system, \$2.083B, is lower than using a concentrated ISRU system, \$2.147B. As we can see in both solutions, the transportation to \ac{GEO} always receives support from the propellant generated by \ac{ISRU}.

\begin{table}[hbt!]
\centering
\caption{Comparison of ISRU systems}
\label{tab:comparison-isru-systems}
\begin{tabular}{lcc}
\hline
\textbf{}                        & \textbf{Concentrated ISRU} & \textbf{Distributed ISRU} \\
\hline
\multicolumn{3}{l}{\textbf{Without Setup Phase}} \\
\hline
Total Mission Cost              
& \$2.922B  &\$2.916B    \\
ISRU Mass Deployed    
& 11,028 kg   & 11,601 kg \\
\hline
\multicolumn{3}{l}{\textbf{With Setup Phase}} \\
\hline
Total Mission Cost             
& \$2.147B & \$2.083B \\
ISRU Mass Deployed              
& 6,300 kg   & 9,295 kg                 \\
\hline
\end{tabular}
\end{table}

This result analysis shows that with a proper setup phase, a distributed \ac{ISRU} system can support the cislunar logistics processes with multiple deployment demands at a lower cost. Note that the conclusion from this case study is based on the specific mission demands and mission operation assumptions. It may not be valid for the performance of distributed \ac{ISRU} in general. However, the main contribution of this paper is to enable such a trade study to compare the performance of distributed \ac{ISRU} and concentrated \ac{ISRU} in supporting a complex multi-mission logistics campaign.

\subsubsection{Sensitivity Analysis} \label{Sensitivity Analysis}

The sensitivity analysis in this study explores the variability of \ac{ISRU} total mission costs under different operational conditions. This analysis explores how changes in key parameters, such as the mission duration, \ac{ISRU} productivity rates, setup phase inclusion, and production efficiency, impact the overall mission cost.

\begin{figure}[hbt!]
\centering
\includegraphics[width=\textwidth]{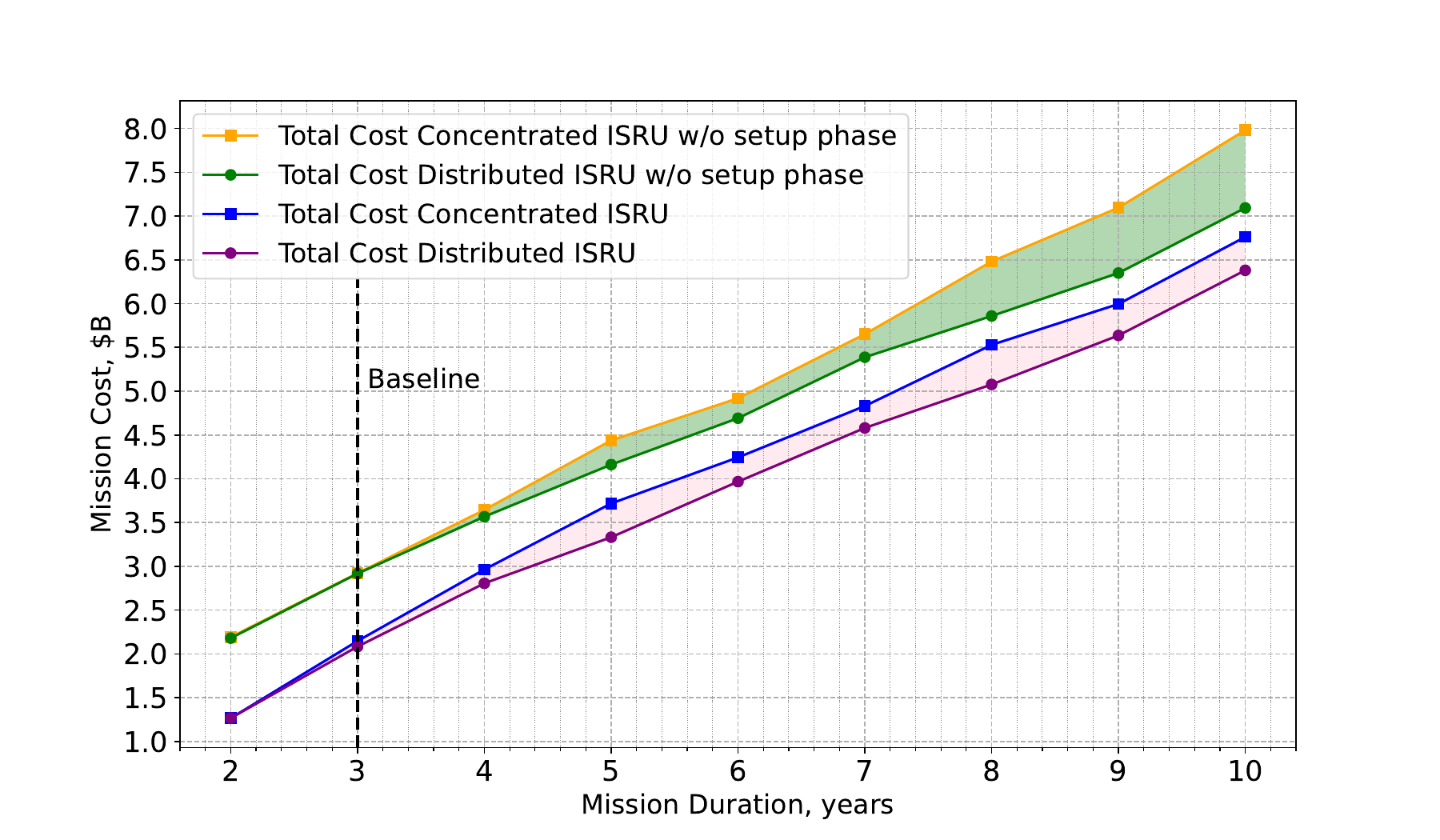}
\caption{Total cost vs mission duration (with/without setup).}
\label{fig:years_total}
\end{figure}

The sensitivity analysis begins by examining the relationship between mission cost and duration for both concentrated and distributed ISRU systems, as shown in Fig. \ref{fig:years_total}. The results indicate that while costs increase with mission duration, distributed ISRU consistently achieves lower costs. The baseline at year 3 marks a turning point where cost trends begin to diverge, highlighting the long-term benefits of distributed \ac{ISRU}'s multi-location deployment strategy.

\begin{figure}[hbt!]
\centering
\includegraphics[width=\textwidth]{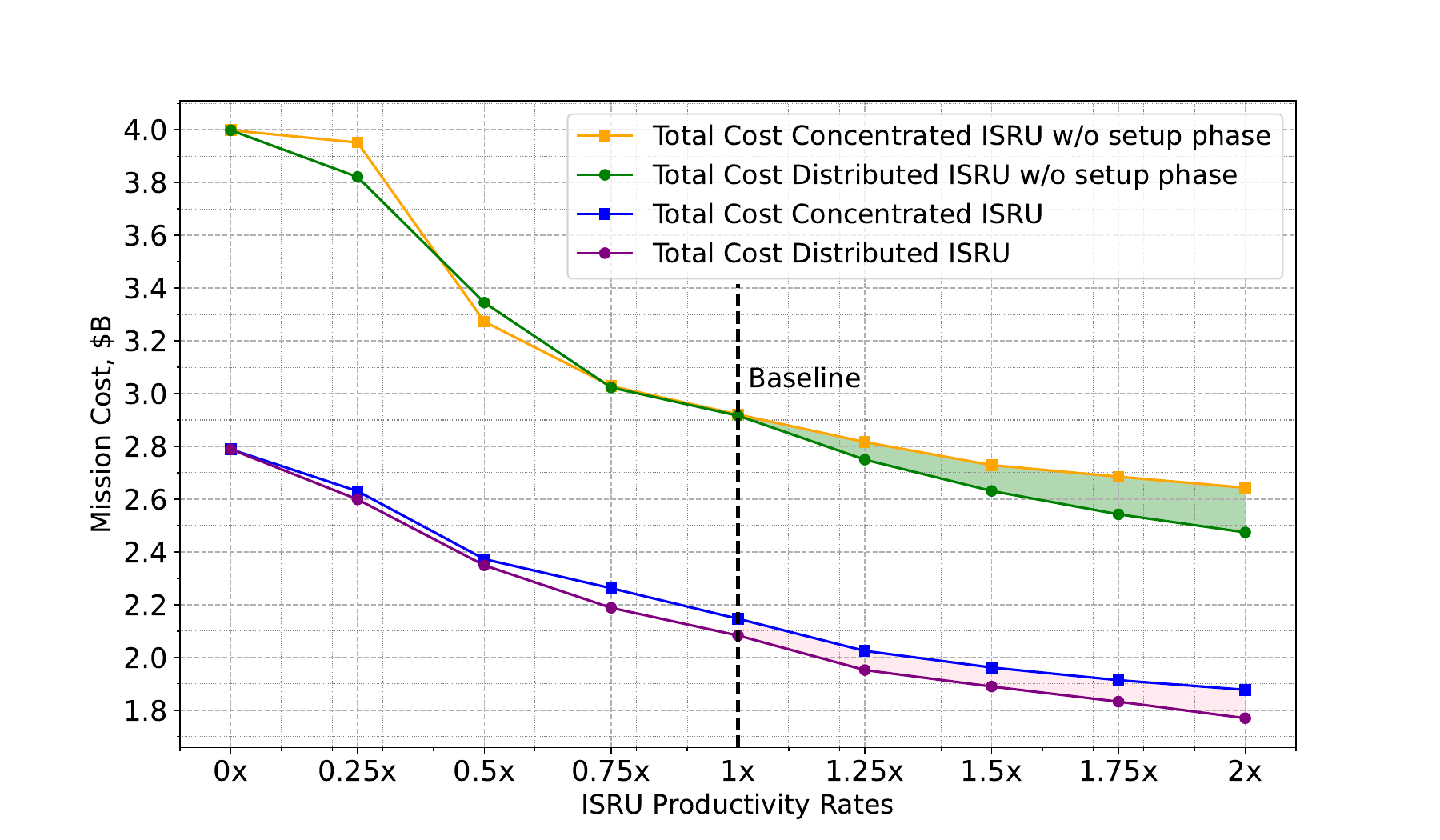}
\caption{Total Cost vs ISRU Productivity Rates (With/Without Setup).}
\label{fig:isruratebase_total}
\end{figure}

Beyond mission duration, another critical factor influencing total mission cost is the productivity rate of ISRU systems, which determines resource extraction and processing efficiency. The baseline productivity rates of the \ac{ISRU} systems, exploring how different levels of resource processing efficiency impact mission costs. The \ac{ISRU} productivity rates of resource utilization for extraction and processing were varied across different scenarios to evaluate their impact on total mission costs. Figure \ref{fig:isruratebase_total} shows the correlation between total mission costs and \ac{ISRU} productivity rates for both concentrated and distributed \ac{ISRU} systems, considering the inclusion and exclusion of the setup phase. We notice that as the productivity rates increase, the mission costs for all \ac{ISRU} systems decrease significantly. Notably, distributed \ac{ISRU} with the setup phase shows a steeper reduction in costs compared to scenarios without the setup phase, highlighting that incorporating the setup phase strategically or optimizing its duration can significantly enhance cost-effectiveness. 

The baseline \ac{ISRU} productivity rate at 1.0\(\times\) marks a point where distributed \ac{ISRU} begins to outperform or closely match concentrated \ac{ISRU} in cost savings. Beyond this baseline, the cost benefits of further productivity increases continue to grow, highlighting even greater potential for cost savings through enhanced \ac{ISRU} efficiency. This analysis shows the strategic advantage of distributed \ac{ISRU} systems in achieving significantly lower costs for large-scale missions, especially when the setup phase is optimized or minimized, allowing for more efficient and cost-effective missions.

\begin{figure}[hbt!]
\centering
\includegraphics[width=\textwidth]{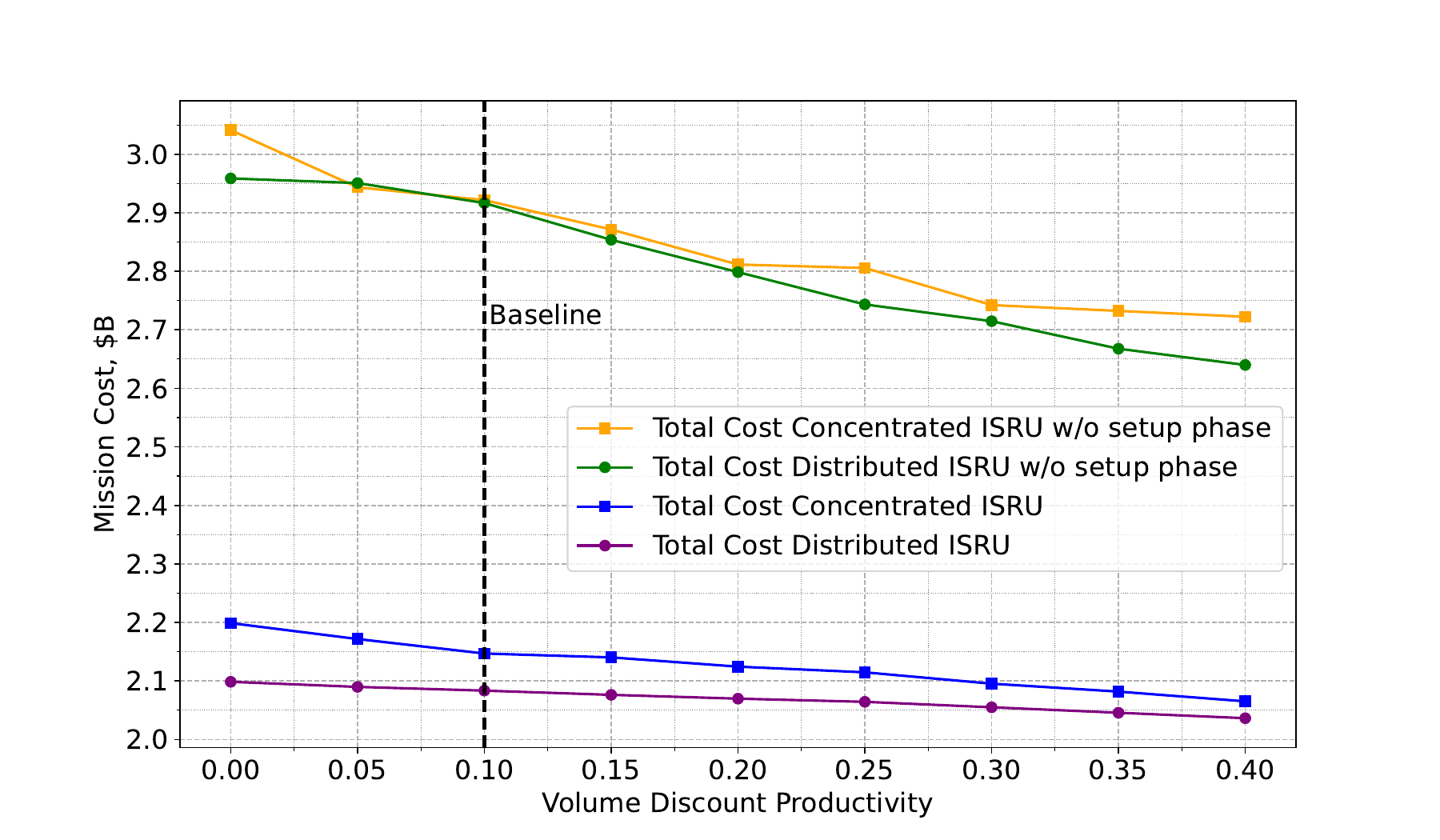}
\caption{Total Cost vs ISRU Volume Discount (With/Without Setup).}
\label{fig:volumediscount_total}
\end{figure}

Following this, the effect of volume discount on productivity was assessed, highlighting how increased production volumes can reduce costs. Figure \ref{fig:volumediscount_total} examines the impact of volume discounts on \ac{ISRU} productivity, indicating that as volume discount of productivity increases, mission costs decrease substantially. The graph demonstrates that while the exclusion of a setup phase initially leads to higher costs, the advantages of volume discounts become increasingly evident in distributed \ac{ISRU} systems as the scale increases.

\begin{figure}[hbt!]
\centering
\includegraphics[width=\textwidth]{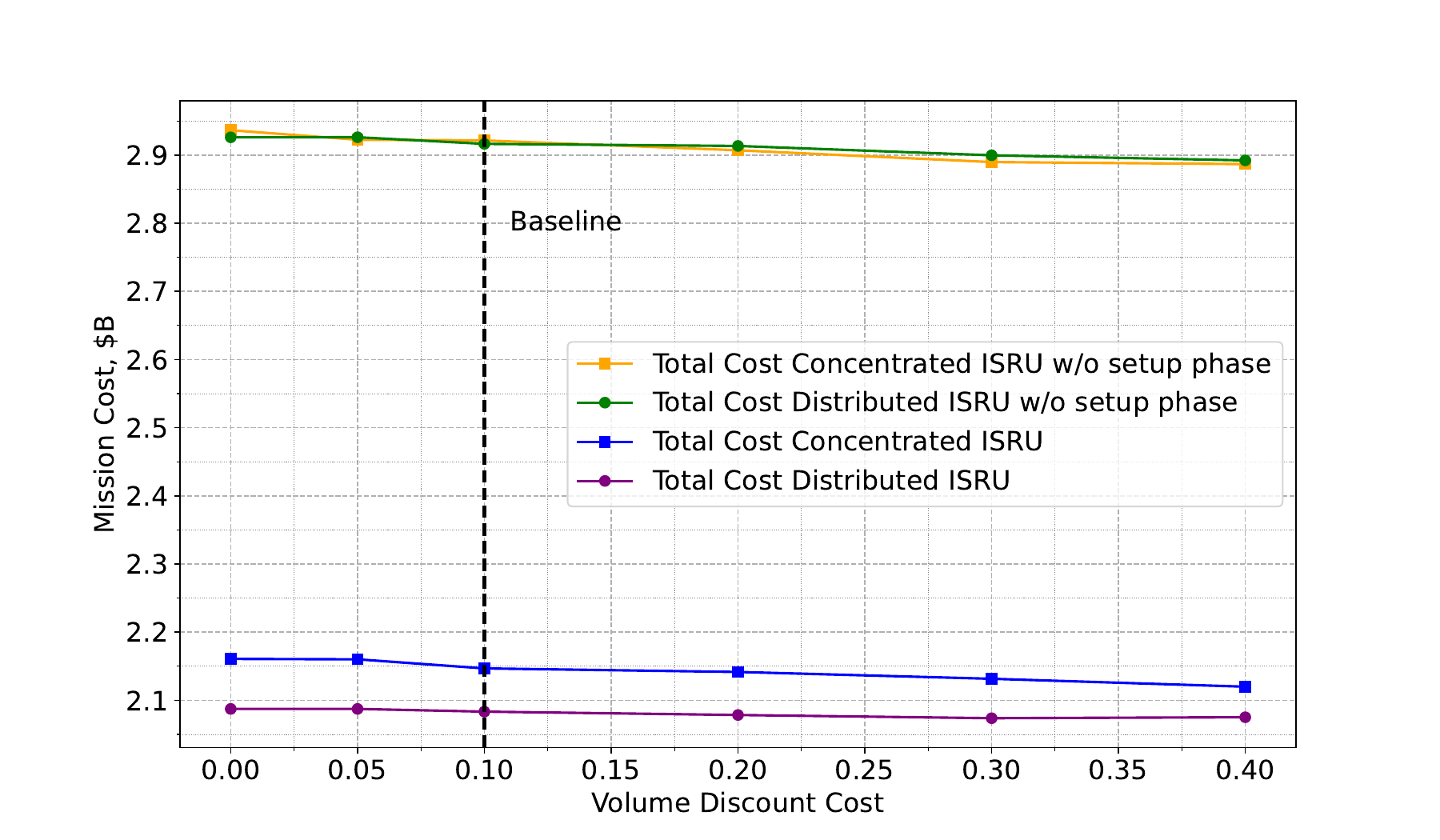}
\caption{Total Cost vs ISRU Cost Discount (With/Without Setup).}
\label{fig:costdiscount_total}
\end{figure}

Next, the impact of cost discounting due to economies of scale is analyzed, and Fig. \ref{fig:costdiscount_total} evaluates the effect of cost discounts on mission expenses across various \ac{ISRU} configurations. The analysis reveals that while cost discounts contribute to lowering mission costs, their impact is relatively modest compared to improvements achieved through productivity rate increases and volume scaling. 

\begin{figure}[hbt!]
\centering
\includegraphics[width=\textwidth]{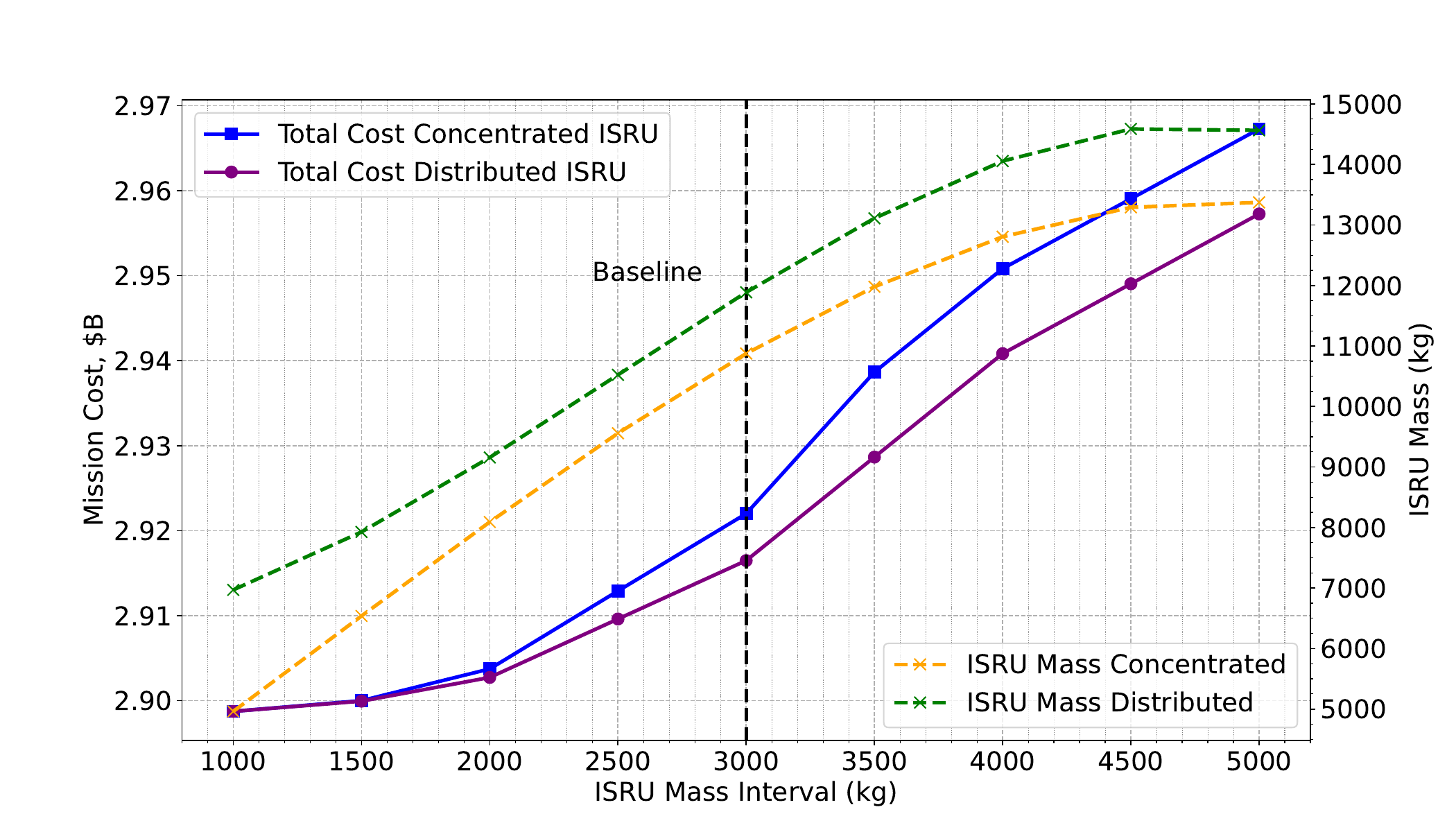}
\caption{Impact of ISRU Mass Scaling on Cost and System Mass without setup phase.}
\label{fig:isrumassinterval}
\end{figure}

Finally, the analysis considered the effect of different mass scaling intervals for the \ac{ISRU} systems, providing insights into how changes in the economies of scale on deployed systems influence overall mission costs. A larger ISRU mass scaling interval means that volume and cost discounts occur at a higher ISRU system mass.
Figure \ref{fig:isrumassinterval} compares the total mission cost and ISRU mass for distributed and concentrated \ac{ISRU} systems across various mass scaling intervals. The concentrated \ac{ISRU}, with lower system mass, has a higher mission cost, which reflects the challenge of deploying a centralized infrastructure entirely on the lunar surface. This approach leads to higher expenses despite the reduced mass. In contrast, the distributed ISRU has a higher system mass, but achieves lower mission costs. This occurs due to the strategic deployment of some subsystems closer to Earth, such as \ac{EML1}, reducing the logistics and transportation costs. This shows that while distributed \ac{ISRU} requires more mass to be deployed in the discussed mission scenarios, it is cost-saving due to the flexibility of multi-location deployment strategy. Moreover, when the ISRU mass scaling interval becomes larger, meaning a weaker economies of scale effect, the distributed \ac{ISRU} deployment strategy leads to a lower mission cost.

This approach offers a deeper understanding of the interplay between ISRU productivity, volume and cost discounts, and mass scaling, ultimately guiding the identification of optimal operating conditions and cost-saving strategies in mission design.

\subsection{Discussion} \label{Discussion}

This study examines the trade-offs between concentrated and distributed ISRU architectures, considering their cost, mass, and logistical complexities. While distributed ISRU reduces reliance on Earth-launched propellant, it introduces operational complexity due to transportation constraints between the lunar surface and \ac{EML1}. 
Unlike a concentrated \ac{ISRU} system, a distributed \ac{ISRU} requires additional logistical coordination, increasing the risk of transportation disruptions and delays in resource delivery. Given that the cost difference between concentrated and distributed ISRU is relatively small and less than 5\%. This is within the range of uncertainty typical for space mission cost estimates, and thus additional flexibility in logistics planning is required to mitigate potential risks.

While distributed ISRU introduces greater complexity, it strengthens mission robustness by providing multiple supply nodes and reducing dependence on Earth-based launches. To further quantify the impact of uncertainties in transportation, operational constraints, and \ac{ISRU} system performance, a sensitivity analysis was conducted, extending the mission horizon beyond three years to determine whether distributed ISRU remains advantageous over longer durations. The results indicate that even as uncertainties are introduced, distributed ISRU continues to demonstrate long-term benefits in resource availability and cost efficiency.

The inherent uncertainties in cost modeling for space hardware must also be acknowledged. Specifically, cost estimates are influenced by technology readiness levels, operational constraints, and economies of scale, which introduce significant variability. Figures \ref{fig:years_total} and \ref{fig:isrumassinterval} further highlight that distributed ISRU remains favorable under assumed scaling relationships, though real-world deployment scenarios may introduce unexpected cost and mass deviations.

\section{Conclusion} \label{Conclusion}
This paper proposes an optimization framework for distributed resource logistics system design. It integrates piecewise linear sizing and cost estimation models developed based on economies of scale into network-based space logistics optimization methods. This framework enables concurrent optimization for \ac{ISRU} technology trade studies, deployment strategies, facility location evaluation, and resource logistics after generation, considering nonlinear sizing and cost models. A cislunar logistics case study is conducted to demonstrate the value of the proposed method. The results show that distributed \ac{ISRU} can support a multi-demand cislunar logistics mission at a lower cost than the concentrated \ac{ISRU} under the assumed mission scenario. This case study also indicates that decision-makers can evaluate potential logistics architectures and \ac{ISRU} technologies using the proposed optimization framework. 

The sensitivity analysis highlights the benefits of distributed \ac{ISRU} systems, showing that enhanced \ac{ISRU} productivity rates and effectively managing the setup phase can significantly reduce mission costs. Enhanced productivity and strategic setup in distributed \ac{ISRU} lead to lower costs than concentrated \ac{ISRU} systems. Economies of scale, including volume and cost discounts, further contribute to cost reductions, underscoring the strategic advantage of distributed \ac{ISRU} for large-scale, long-duration missions and optimizing overall cost efficiency. 

Future research directions include assessing the resilience of distributed and concentrated \ac{ISRU} systems under uncertainties and analyzing the impact of these uncertainties on the transportation system. Research can also be done to consider the interactions among different \ac{ISRU} plants in the cislunar system, further refining the optimization of resource logistics for space missions.

\bibliography{sample}

\end{document}